\documentclass[12pt]{article}

\usepackage{amsfonts}
\usepackage{amsmath}
\usepackage{amssymb}

\newtheorem{theorem}{Theorem}
\newtheorem{lemma}[theorem]{Lemma}
\newtheorem{corollary}[theorem]{Corollary}

\newtheorem{definition}[theorem]{Definition}

\newcommand{\R}{\mathbb{R}}
\newcommand{\Q}{\mathbb{Q}}
\newcommand{\C}{\mathbb{C}}
\newcommand{\Pf}{\mathbb{F}_1}
\newcommand{\N}{\mathbb{N}}
\newcommand{\Z}{\mathbb{Z}}
\newcommand{\CP}{\mathbb{C}\mathcal{P}^2}
\renewcommand{\H}{\mathcal{H}}
\newcommand{\B}{\mathcal{B}}
\newcommand{\A}{\mathcal{A}}
\newcommand{\lra}{\longrightarrow}
\newcommand{\vp}{\varphi}
\newcommand{\logplus}{\stackrel{+}\log}
\newcommand{\Ce}{\mathcal{C}}
\newcommand{\Ra}{\mathcal{R}}
\newcommand{\NN}{\mathcal{N}}
\newcommand{\E}{\mathit{E}}

\newcommand{\nn}{\nonumber}

\begin{document}
\begin{center}
{\LARGE\bf
Height growth of solutions and\vskip 3mm a discrete Painlev\'e equation}

\vskip 6 mm

{\large A Al-Ghassani\footnote{Department of Mathematics and Statistics,
College of Science,
Sultan Qaboos University,
Al-Khodh P.O. Box 36, P.C 123,
Sultanate of Oman} and 
R G Halburd\footnote{Department of Mathematics,
University College London, Gower Street, London WC1E 6BT, UK }
}
\end{center}

\vskip 8 mm

\begin{abstract}
\noindent
Consider the discrete equation
$$
y_{n+1}+y_{n-1}=\frac{a_n+b_ny_n+c_ny_n^2}{1-y_n^2},
$$
where the right side is of degree two in $y_n$ and where the coefficients $a_n$, $b_n$ and $c_n$ are rational functions of $n$ with rational coefficients.   Suppose that there is a solution such that
for all sufficiently large $n$, $y_n\in\mathbb{Q}$ and the height of 
$y_n$ dominates the height of the coefficient functions $a_n$, $b_n$ and $c_n$.  We show that if the logarithmic height of $y_n$
grows no faster than a power of $n$ then either the equation is a well known discrete Painlev\'e equation
${\rm dP}_{\!\rm II}$  or its autonomous version or  $y_n$ is also an admissible solution of a discrete Riccati equation.
This provides further evidence  that slow height growth is a good detector of integrability.% ({\em Diophantine integrability}).
\end{abstract}
%\ams{39A12, 11Z05, 37J35.}
%\pacs{???,???}

%\maketitle

%\pagestyle{headings}
\section{Introduction}

For discrete equations, integrability appears to be closely related to the slow growth of various measures of complexity %(Veselov 1992)
\cite{veselov}.  For example,  algebraic entropy %(Falqui \& Viallet 1993, Hietarinta \& Viallet 1998, Bellon \& Viallet 1999)
\cite{FaVi93,HiVi98,alg-entropy} 
measures the degree growth of iterates as a function of the initial conditions, while in the Nevanlinna approach 
\cite{AHH,HalburdK-JPA},
%(Ablowitz et al 2000; Halburd \& Korhonen 2007) 
one considers the order of growth of meromorphic solutions.  A discrete equation on a number field is said to be {\em Diophantine integrable} 
if the logarithmic height of solutions grows polynomially \cite{halburd:05}.
% (Halburd 2005).
The last two approaches are connected by Vojta's dictionary \cite{vojta},
%(Vojta 1987), 
which relates ideas from Nevanlinna theory (the value distribution of meromorphic functions) to those in Diophantine approximation.
See \cite{ghrv-jpa} for a survey of different approaches to detecting integrability in discrete systems.

This paper concerns a Diophantine analogue of a classification result of Halburd and Korhonen %(2007a)
\cite{HalburdK-LMS} 
using Nevanlinna theory.  Specifically, we will study discrete
equations of the form
\begin{equation}
\label{main eqn}
y_{n+1}+y_{n-1}=\frac{a_n+b_ny_n+c_ny_n^2}{1-y_n^2},
\end{equation}
where $a_n$, $b_n$ and $c_n$ are in $\mathbb{Q}(n)$ and the degree of the right side of equation (\ref{main eqn}) is two.
The logarithmic height of a rational number $x=a/b$, where $a$ and $b$ have no common factors, is $h(x)=\log H(x)$, where $H(x)=\max\{|a|,|b|\}$ is the height.
A discrete equation such as (\ref{main eqn}) is said to be {\em Diophantine integrable} if the logarithmic height of its solution $y_n$ over a number field grows no faster than a power of $n$ \cite{halburd:05}.
%(Halburd 2005).
Abarenkova {\it et al} \cite{abarenkovaabhm:99} used height growth to estimate the entropy of a map.  Slow height growth has been used as an efficient numerical test in
\cite{hone:06,AnMaVi05,hones:08,honep:09,fordym}.
% Hone 2006, Angl\`es d'Auriac {\it et al} \cite{AnMaVi05}, Hone and Swart 2008, Hone and Petrera 2009 and Fordy and Marsh 2011.

The purpose of this paper is to prove the following.
\begin{theorem}
%\begin{main result}
\label{mainthm}
Let $r_0$ be sufficiently large and 
let $(y_n)_{n\ge r_0}\subset\mathbb{Q}\setminus\{-1,1\}$ be a solution of (\ref{main eqn}),
where $a_n$, $b_n$ and $c_n$ are rational functions of $n$ with coefficients in $\mathbb{Q}$ and the right  side of (\ref{main eqn}) is of degree two in $y_n$. 
If
\begin{equation}
\sum_{n=r_0}^r \max\{1,h(a_n),h(b_n),h(c_n)\}
=
o\left(
\sum_{n=r_0}^r h(y_n)
\right)
\label{admissible}
\end{equation}
as $r\to\infty$, then either
\begin{enumerate}
\item $a_n=\alpha n+\beta$, $b_n=\gamma$, $c_n=0$ for constants $\alpha, \beta, \gamma$; or
\item $y_n$ also solves the discrete Riccati equation
\begin{equation}
\label{ourriccati}
y_{n+1}=\frac{1/2(a_n+\theta b_n-2\theta)+y_n}{1-\theta y_n}, \mbox{  where $\theta=-1$ or 1; or}
\end{equation}
\item$\displaystyle
%\underset{r\longrightarrow \infty}
\limsup_{r\to\infty} \frac{\log\log \sum_{n=r_0}^rh(y_n)}{\log r}\geq 1$.
\end{enumerate}
\end{theorem}
%\end{main result}
This result first appeared in the PhD thesis of the first author.

A solution of equation (\ref{main eqn}) satisfying (\ref{admissible}) will be called {\em admissible}.
Of the three possible outcomes described in Theorem \ref{mainthm}, the first says that equation (\ref{main eqn}) is the discrete Painlev\'e equation ${\rm dP}_{\rm II}$ 
(see Nijhoff and Papageorgiou \cite{nijhoffp})
or its autonomous version ($\alpha=0$), the second says that $y_n$ solves a well known linearisable discrete Riccati equation and the third implies that $h(y_n)$ grows faster than any power of $n$. If equation (\ref{main eqn}) has more than two one-parameter families of admissible solutions, they cannot both solve discrete Riccati equations of the form described by the second conclusion unless the equation is ${\rm dP}_{\rm II}$.
In case 1 with $\alpha=0$, equation (\ref{main eqn}) can be derived from the addition law on an elliptic curve, for which it is known that the logarithmic height grows quadratically.
% Hence, the theorem says that if for all admissible solutions $h(y_n)$ grows polynomially, equation (\ref{main eqn}) is the discrete analogue of the second Painlev\'e equation. 

The full version of ${\rm dP}_{\rm II}$ allows $a_n$ to have the more general form $a_n=\alpha n+\beta+\delta(-1)^n$, where $\delta$ is another constant.  We do not capture this form as we have assumed that the coefficients $a_n$, $b_n$ and $c_n$ are rational functions of $n$.  This assumption simplifies some of the arguments.

Diophantine integrability is a property of all solutions, not just those that are admissible. Our method involves working with one solution at a time, so an admissibility-type condition is necessary to avoid counterexamples in which $a_n$, $b_n$ and $y_n$ are chosen arbitrarily and then $c_n$ is determined by equation (\ref{main eqn}).
%{\bf singularity confinement} \cite{sc}

Of central importance in our proof of Theorem \ref{mainthm} is the fact that there is a simple relationship  between the height of a rational number $x$ and a certain sum over all non-trivial absolute values of $x$. For a fixed prime $p$, the $p$-adic absolute value of a non-zero rational number $x$ is $|x|_p=p^{-r}$, where $x=\frac mn p^r$ for integers $m$, $n$ and $r$ such that $p\!\!\not |\, mn$.  The $p$-adic absolute values are non-Archimedean, which means that they satisfy the stronger triangle inequality
$|x+y|_p\le\max\{|x|_p,|y|_p\}$.  The usual absolute value, denoted by $|\,\cdot\,|_\infty$, is Archimedean.  Ostrovski's Theorem says that, up to equivalence, the only non-trivial
absolute values on $\mathbb{Q}$ are the $p$-adic absolute values, $|\,\cdot\,|_p$ and the usual absolute value $|\,\cdot\,|_\infty$.  In terms of these absolute values, we have the important identity
\begin{equation}
\label{logh}
h(x)=\sum_{p\le\infty} \log^+|x|_p,
\end{equation}
where the sum is taken over all finite primes and $p=\infty$ (the ``prime at infinity'') and $\log^+y:=\max\{0,\log y\}$.

%{\bf generalisations to number fields}

One of the first properties of discrete equations to be used to identify discrete Painlev\'e equations was singularity confinement \cite{grammaticosrp,ramanigh}, which involves the behaviour of solutions as one iterates through a singularity of the equation.  For equation (\ref{main eqn}), one needs to examine the singular values $y=1$ and $y=-1$.  In order to resolve indeterminacies that arise in future iterates, we consider the initial conditions $y_{k-1}=\kappa$, $y_k=\theta+\epsilon$, where $\kappa$ is arbitrary, $\theta^2=1$ and $\epsilon$ is small (as we will take the limit $\epsilon\to 0$ after a finite number of steps).
Generically we find that, after taking the limit $\epsilon\to 0$, $y_m=\infty$ for infinitely many $m>k$.  However, for certain choices of $a_n$, $b_n$ and $c_n$, the singularity appears to be confined to a finite number of iterates.

At the heart of the proof of Theorem \ref{mainthm} are some calculations that look very much like those described above for singularity confinement.  The main difference is that we have to consider not only the case in which $1-\theta y_{k}$ is small with respect to the usual absolute value, but also cases in which it is small with respect to $p$-adic absolute values.
The identity (\ref{logh}) is eventually used to convert certain statements about absolute values into statements about logarithmic heights.
% A key step on our way to proving Theorem \ref{mainthm} is %
%For example, in the following, in which f
The following theorem is an example of 
such an expression of singularity (non-)confinement in terms of absolute values. It should be stressed, however, that we do not make assumptions about the long term behaviour of solutions or whether they are eventually confined. 
For each absolute value $|\cdot |_p$, $\epsilon_k\equiv \epsilon_{k,p}$, which is defined precisely in equation (\ref{eps2}), determines a length scale in terms of the coefficients
 $a_k$, $b_k$, $c_k$ and a finite number of their shifts. 
\begin{theorem}
%\begin{main2}
\label{SC}
\textit{Let $(y_n)_{n=k-1}^{k+3} \subset \mathbb{Q}\setminus \{-1,1\}$ satisfy}
%\begin{equation*}
\begin{equation*}
y_{n+1}+y_{n-1}=\frac{a_n+b_ny_n}{1-y_n^2},
\end{equation*}
%\end{equation*}
where $k$ is sufficiently large and the right hand side of the equation is irreducible. Assume that for a fixed absolute value $|.|_p$ ($p\leq \infty$) we have $|y_{k-1}|_p\leq |1-\theta y_k|_p^{-1/2}$ for $\theta=1$ or $-1$. Furthermore, for sufficiently small $\delta>0$, if $|1-\theta y_k|_p<\epsilon_k$ (where $\epsilon_k$ is defined in (\ref{eps2})), then 
\begin{enumerate}
\item[(i)] $y_{k+1}=\frac{\textstyle a_k+\theta b_k}{\textstyle2(1-\theta y_k)}+A_k$,
where $|A_k|_p\leq |1-\theta y_k|_p^{-1/2}$ if $p<\infty$ and if $p=\infty$, $|A_k|_\infty\leq \frac{11}{10}\cdot |1-\theta y_k|_\infty^{-1/2}$.
\item[(ii)] $y_{k+2}=-\theta+\left(\frac{\textstyle \theta a_k+b_k-2b_{k+1}}{\textstyle a_k+\theta b_k}\right)(1-\theta y_k)+B_k$, \\
where $|B_k|_p\leq |1-\theta y_k|_p^{3/2-5\delta}$ if $p<\infty$ and if $p=\infty$, $|B_k|_\infty\leq \frac{1}{2}\cdot |1-\theta y_k|_\infty^{3/2-5\delta}$.
\item[(iii)] $y_{k+3}=\frac{\left(\textstyle a_{k+2}-\theta b_{k+2}-\theta (\theta a_k+b_k-2b_{k+1})\right)}{\textstyle 2(1+\theta y_{k+2})}+C_k$,
 where $|C_k|_p\leq |1+\theta y_{k+2}|_p^{-(2/3+2\delta)}$ if $p<\infty$ and if $p=\infty$, $|C_k|_\infty\leq 2|1+\theta y_{k+2}|_\infty^{-(2/3+2\delta)}$.
\end{enumerate}
\end{theorem}
Theorems \ref{mainthm} and \ref{SC} can easily be extended to arbitrary number fields (finite field extensions of the rationals) as there is a simple analogue of the identity (\ref{logh}) in this case.

In \cite{halburd:05}, it was shown that if an equation of the form
\begin{equation}
\label{clas}
y_{n+1}+y_{n-1}=R(n,y_n),
\end{equation}
where $R$ is rational, has an admissible solution, then $\mbox{deg}_yR(n,y)\le 2$.  The case $R(n,y_n)=(a_n+b_ny_n+c_ny_n^2)/y_n^2$ was studied in 
%Here, $R$ is rational in $y_n$ with coefficients that are rational functions in $n$ and rational numbers; also the degree of $R\leq 2$. 
\cite{HalburdM}.
% studied in detail a particular case of the class of equations (\ref{clas})(namely $y_{n+1}+y_{n-1}=\frac{\alpha_n+\beta_ny_n+\gamma_ny_n^2}{y_n^2}$).  Here we analyse a different case of the same class of equations (\ref{clas}). However, t
There are essential technical difficulties which distinguish  the two cases and consequently the analysis used to treat each of them. 
The fact that, for certain $a_n$, $b_n$ and $c_n$, there are solutions of equation (\ref{main eqn})  that also satisfy discrete Riccati equations requires a more subtle analysis.
%In the equation considered by Morgan in his PhD thesis~\cite{halburd5}, there is one singularity at $y_n=0$ of multiplicity 2. 
%In  equation (\ref{main eqn}) we are considering, we have 2 distinct singularities, each of multiplicity 1 at $y_n=1$ and $y_n=-1$. Unlike the equation considered by Morgan, we have a major technical difficulty that arose in our case, in which for certain forms of the equation coefficients, $y_n$ solves a discrete Riccati equation.   

%%%%%%%%%%%%%%%%%%%%%%%%%% subsection 4.1 %%%%%%%%%%%%%%%%%%%%%%%%%%%%%%%%%%%%%%%%%%%%%%%%%%
\section{Proof of Theorem \ref{mainthm}}

%\newtheorem{def5}{Definition}[section]
%\begin{def5}
%\textit{Let $(y_n)\subset\mathbb{Q}$ be a solution for some discrete equation. We define the \end{def5}

First we consider the case in which $c_n\not\equiv -2$, $0$  or $2$.
We introduce a quantity $\epsilon_n$, which 
provides a scale with respect to which we measure distances between iterates and certain singular values.
%serves as our measure for the size of the iterate $y_n$ in equation (\ref{main eqn}) and to measure its distance from the singularities $\pm1$.  
For any finite set of rational functions $\{f_1,\ldots, f_m\}$ of $n$, there exists $K\in\mathbb{N}$ such that for every function $f_j$ that is not identically zero,
$f_j(n)$ is finite and nonzero for all $n\ge K$.  Throughout this paper we will refer to such an integer $K$, which may need to be increased a finite number of times, without further comment.
 Since the right side of equation (\ref{main eqn}) is of degree 2, neither of the rational functions $a_n+ b_n+c_n$ nor $a_n- b_n+c_n$ vanishes identically.
 For $n>K$, define the sets
$X_{n,0} =\Big\{1/2,b_n,c_n,c_{n-1}^{-1},c_{n+1}^{-1}\Big\}$,
$ X_{n,\pm} =\{
 (a_n\pm b_n+c_n)^{-1},(a_{n+1}\pm b_{n+1}+c_{n+1})/2,(a_{n-1}\pm b_{n-1}+c_{n-1})/2,
  (c_{n+1}\pm 2)^{-1},(c_{n-1}\pm 2)^{-1}
 \}$ and let $X_n=X_{n,+}\cup X_{n,0}\cup X_{n,-}$.
For a fixed sufficiently small $\delta>0$ we define $\epsilon_n$ ($\forall n>K$) by
\begin{equation}
\label{defeps}
 \epsilon^{-\delta}_n=\kappa_p\max_{x\in X_n}\{|x|_p\},
\end{equation}
 where $\kappa_p=1$ for $p< \infty$ and $\kappa_\infty=10$.  
 %Although the purpose of any $\epsilon_n$ is the same, we introduce different definitions to avoid technical issues which might arise in the proof of lemmas and theorems. 
Equation (\ref{defeps}) allows us to estimate certain combinations of the coefficients $a_n$, $b_n$, $c_n$ in terms of $\epsilon_n$.
For example, for $p<\infty$, we have $|c_n|_p\leq \epsilon_n^{-\delta}\Rightarrow \epsilon_n^{\delta}\leq |c_n|_p^{-1}$. For the Archimedean absolute value ($p=\infty$), we have $10|c_n|_{\infty}\leq \epsilon_n^{-\delta}\Rightarrow \epsilon_n^{\delta}\leq \frac{1}{10}|c_n|_{\infty}^{-1}$.   

%%%%%%%%%%%%%%%% presenting the lemmas & theorems%%%%%%%%%%%%%%%%%
%\newtheorem{lem4}{Lemma }[section]
\begin{lemma}
%\begin{lem4} 
\textit{Let $(y_n) \subset \mathbb{Q}\setminus\{-1,1\}$ be a solution of equation (\ref{main eqn})
% \begin{equation*}
%y_{n+1}+y_{n-1}=\frac{a_n+b_ny_n+c_ny_n^2}{1-y_n^2},
%\end{equation*}
where $a_n, b_n$ and $c_n$ are in $\mathbb{Q}(n)$ and $c_n$ is a rational function not identically $0$ or $\pm2$. Furthermore, assume that the numerator and the denominator of $(\ref{main eqn})$ are coprime. For a fixed prime $p\leq \infty$ and $k>K$, let $\epsilon_k$ be as defined in $(\ref{defeps})$. If $|1-\theta y_k|_p<\epsilon_k$ for $\theta =1$ or $-1$, then either
%\begin{equation*}
$$
|y_{k+1}|_p\geq \frac{1}{|1-\theta y_k|^{1-\delta}_p}  \hspace{0.3cm} \mbox{and} \hspace{0.3cm} |1\pm\theta y_{k+2}|_p \geq \epsilon_{k+2},
$$
%\end{equation*}
or
%\begin{equation*}
$$
|y_{k-1}|_p\geq \frac{1}{|1-\theta y_k|_p^{1-\delta}}  \hspace{0.3cm} \mbox{and} \hspace{0.3cm} |1\pm\theta y_{k-2}|_p \geq \epsilon_{k-2}.
$$
%\end{equation*}
}
\label{lemma3.1.1}
\end{lemma}
%\end{lem4}
%%%%%%%%%%%%%%%%%%%%%%%%%% Proof of lemma3.1.1 %%%%%%%%%%%
%\begin{description}

\noindent{\bf Proof :}
% First note that $a_n\pm b_n+c_n$ are rational functions not identically zero for all $n$, otherwise, it contradicts the assumption that the denominator and the numerator are coprime.
The definition of $\epsilon_n^{-\delta}$ in (\ref{defeps}) implies that $\epsilon_n\le 1$. Furthermore when $p=\infty$ or $p=2$,  $\epsilon_n< 1$.
% when $p<\infty$ and $p\neq 2$. 
First we consider the non-Archimedean case for a fixed prime $p<\infty$.
Let $|1-\theta y_k|_p<\epsilon_k$ for some $k>K$, where $\theta =1$ or $-1$. 
From equation (\ref{main eqn}) we have
 \begin{equation}
\label{cross1}
(y_{k+1}+y_{k-1})(1+\theta y_k)=\frac{a_k+\theta b_k+c_k}{1-\theta y_k}-\theta b_k-c_k(1+\theta y_k).
\end{equation}
So from equations (\ref{defeps}) and (\ref{cross1}),
 \begin{eqnarray}
\label{ubound}
&|1-\theta y_k|_p^{-(1-\delta)}%&=&\frac{|1-\theta y_k|_p^\delta}{|1-\theta y_k|_p}
<\frac{\epsilon^\delta_k}{|1-\theta y_k|_p}% ,\hspace{0.3cm}\mbox{(from $|1-\theta y_k|_p<\epsilon_k$)}\nonumber\\
\leq \frac{|a_k+\theta b_k+c_k|_p}{|1-\theta y_k|_p}\nonumber\\
\leq&|(y_{k+1}+y_{k-1})(1+\theta y_k)+\theta b_k+c_k(1+\theta y_k)|_p\nonumber\\
\leq&\max\{|y_{k+1}+y_{k-1}|_p\cdot|1+\theta y_k|_p,|b_k|_p,|c_k|_p\cdot|1+\theta y_k|_p\}.
\end{eqnarray}
From 
(\ref{defeps}), % we have
%\begin{equation*}
$
|b_k|_p\leq\epsilon^{-\delta}_k\leq\epsilon^{-(1-\delta)}_k<|1-\theta y_k|_p^{-(1-\delta)}$.
%\end{equation*}
Similarly, $|c_k|_p<|1-\theta y_k|_p^{-(1-\delta)}$. 
Also, 
$|1+\theta y_k|_p=|2-(1-\theta y_k)|_p\leq \max\{|2|_p,|1-\theta y_k|_p\}\leq \max\{1,\epsilon_k\}=1$.
Therefore, %the inequality in 
(\ref{ubound}) reduces to
$|1-\theta y_k|_p^{-(1-\delta)}%& \leq&%\max\{|y_{k+1}+y_{k-1}|_p\cdot|1+\theta y_k|_p,|c_k|_p\cdot|1+\theta y_k|_p\}\nonumber\\
%&=&|1+\theta y_k|_p.\max\{|y_{k+1}+y_{k-1}|_p,|c_k|_p\}\nonumber\\
\leq %\max\{|y_{k+1}+y_{k-1}|_p, |c_k|_p\}=
|y_{k+1}+y_{k-1}|_p
\leq  \max\{|y_{k+1}|_p,|y_{k-1}|_p\}$.
%Therefore, either the maximum is $|y_{k+1}|_p$ where we would have the proof completed for $|y_{k+1}|_p\geq \frac{1}{|1-\theta y_k|_p^{(1-\delta)}}$ or the maximum is $|y_{k-1}|_p$ where in this case we would complete the proof of $|y_{k-1}|_p\geq \frac{1}{|1-\theta y_k|_p^{(1-\delta)}}$. 
Without loss of generality, we choose the maximum to be $|y_{k+1}|_p$ and for the rest of the proof we use $|y_{k+1}|_p\geq |1-\theta y_k|_p^{-(1-\delta)} $.\\

%In order to show that $|1\pm \theta y_{k+2}|_p\geq \epsilon_{k+2}$, we first show that $|y_{k+2}+\theta+c_{k+1}|_p$ is small. 
%This means that $y_{k+2}$ is $p$-adically close to a rational number different from $\pm 1$ since by assumption $c_{k+1}$ is not identically $0,\pm 2$ .  
From $|1-\theta y_k|_p<\epsilon_k$ and $|y_{k+1}|_p\geq {|1-\theta y_k|_p^{-(1-\delta)}}$, we have
%\begin{equation*}
$$
{\epsilon_k^{-(1-\delta)}}<%\frac{1}{|1-\theta y_k|_p^{1-\delta}}\leq 
|y_{k+1}|_p%=|1-(1-y_{k+1})|_p
\leq \max \{1,|1\pm y_{k+1}|_p\}=|1\pm y_{k+1}|_p.
$$
%\end{equation*}
Rewriting equation (\ref{main eqn}) as
% \begin{equation*}
%y_{k+2}+y_k=\frac{a_{k+1}+b_{k+1}y_{k+1}-c_{k+1}(1-y_{k+1}^2)+c_{k+1}}{1-y_{k+1}^2}.
%\end{equation*}
%Adding $\theta +c_{k+1}-y_k$ to both sides of this equation and using partial fractions, we get the following:
%\begin{equation}
%\label{terms}
$y_{k+2}+c_{k+1}+\theta=(a_{k+1}+b_{k+1}+c_{k+1})/[2(1-y_{k+1})]+(a_{k+1}-b_{k+1}+c_{k+1})/[2(1+y_{k+1})]+\theta(1-\theta y_k)$,
%\end{equation}
we have
%Therefore, $|1-y_{k+1}|_p^{-1}<\epsilon_k^{1-\delta}$and similarly $|1+y_{k+1}|_p^{-1}<\epsilon_k^{1-\delta}$.
%Applying the $p$-adic absolute value ($p<\infty$) to both sides of (\ref{terms}) we have
%\begin{eqnarray}
%\label{ck+1}
$|y_{k+2}+c_{k+1}+\theta|_p
%\leq \max \Bigg\{ \frac{|\frac{1}{2}|_p.|a_{k+1}+b_{k+1}+c_{k+1}|_p}{|1-y_{k+1}|_p},\nonumber\\
% &\quad \frac{|\frac{1}{2}|_p.|a_{k+1}-b_{k+1}+c_{k+1}|_p}{|1+y_{k+1}|_p},|1-\theta y_k|_p\Bigg\} <%\max\{\epsilon_k^{1-2\delta},\epsilon_k^{1-2\delta},\epsilon_k\}=
<\epsilon_k^{1-2\delta}$.
%\end{eqnarray}
%The maximum is $\epsilon_k^{1-2\delta}$. since $\epsilon_k\leq1$ and $0<\delta<\frac{1}{2}$. 
From  (\ref{defeps}),
\begin{eqnarray*}
\epsilon_k &\leq& \epsilon_k^\delta\le |c_{k+1}+\theta\pm\theta|_p=|(y_{k+2}+c_{k+1}+\theta)\pm\theta(1\mp\theta y_{k+2})|_p\nonumber\\
&\leq& \max \{|y_{k+2}+c_{k+1}+\theta|_p,|1\mp\theta y_{k+2}|_p\}=|1\mp\theta y_{k+2}|_p,
\end{eqnarray*}
%\begin{eqnarray}
%\label{epsk}
%\epsilon_k^\delta &\leq& |c_{k+1}|_p=|(y_{k+2}+c_{k+1}+\theta)-\theta(1+\theta y_{k+2})|_p\nonumber\\
%&\leq& \max \{|y_{k+2}+c_{k+1}+\theta|_p,|1+\theta y_{k+2}|_p\}.
%\end{eqnarray}
for $\delta<1/3$,
which proves the lemma for non-Archimedean absolute values ($p<\infty$).
% ($\forall p<\infty$). 
%By symmetry, had we considered the case $|y_{k-1}|_p\geq |1-\theta y_k|_p^{-(1-\delta)}$, then we would have obtained $|1\pm \theta y_{k-2}|_p\geq \epsilon_{k-2}$. \\

%\pagebreak

%%%%%%%%%%%%%%%%%%%%%%Archimedean case%%%%%%%%%%%%%%%%%

The Archimedean case ($p=\infty$) is similar. 
%Assume that $|1-\theta y_k|_{\infty}<\epsilon_k$. %As in the non-Archimedean absolute value case, we start from (\ref{cross1}).
%Then
We have
\begin{eqnarray}
\label{archestimate}
&&10|1-\theta y_k|_{\infty}^{-(1-\delta)}
%&=&\frac{10|1-\theta y_k|_{\infty}^{\delta}}{|1-\theta y_k|_{\infty}},\nonumber\\
<\frac{10\epsilon_k^\delta}{|1-\theta y_k|_{\infty}}%\hspace{0.3cm} \mbox{(from assumption $|1-\theta y_k|_{\infty}<\epsilon_k$)}
\leq\frac{|a_k+\theta b_k+c_k|_{\infty}}{|1-\theta y_k|_{\infty}}\nonumber\\
&\leq&|(y_{k+1}+y_{k-1})(1+\theta y_k)+\theta b_k+c_k(1+\theta y_k)|_{\infty}\nonumber\nonumber\\
&\leq&|y_{k+1}+y_{k-1}|_{\infty}\cdot|1+\theta y_k|_{\infty}+|b_k|_{\infty}+|c_k|_{\infty}\cdot|1+\theta y_k|_{\infty}.
\end{eqnarray}
Also
$ |1+\theta y_k|_{\infty}\leq |1-\theta y_k|_{\infty}+|2|_{\infty}<\epsilon_k+2<3$.
Finally we have from (\ref{defeps}) that $|b_k|_{\infty}<|1-\theta y_k|_{\infty}^{-(1-\delta)}$ and $|c_k|_{\infty}<|1-\theta y_k|_{\infty}^{-(1-\delta)}$. 
So (\ref{archestimate}) gives
$10|1-\theta y_k|_{\infty}^{-(1-\delta)}<3|y_{k+1}+y_{k-1}|_{\infty}+|b_k|_{\infty}+3|c_k|_{\infty}
%&<&3|y_{k+1}+y_{k-1}|_{\infty}+|1-\theta y_k|_{\infty}^{-(1-\delta)}+3|1-\theta y_k|_{\infty}^{-(1-\delta)},\nonumber\\
<3|y_{k+1}+y_{k-1}|_{\infty}+4|1-\theta y_k|_{\infty}^{-(1-\delta)}$.
Therefore, $2|1-\theta y_k|_{\infty}^{-(1-\delta)}\leq|y_{k+1}+y_{k-1}|_{\infty}\leq |y_{k+1}|_{\infty}+|y_{k-1}|_{\infty}.$
Hence, either $|y_{k+1}|_{\infty}\geq |1-\theta y_k|_{\infty}^{-(1-\delta)}$ or $|y_{k-1}|_{\infty}\geq |1-\theta y_k|_{\infty}^{-(1-\delta)}$, which proves the first assertion of the lemma. Without loss of generality, we take $|y_{k+1}|_{\infty}\geq |1-\theta y_k|_{\infty}^{-(1-\delta)}$. 

%Now we prove that $|1\pm \theta y_{k+2}|_{\infty}\geq \epsilon_{k+2}$. 
%As before, to get the term $y_{k+2}+c_{k+1}+\theta$, we rewrite equation (\ref{main eqn}) in the form (\ref{terms}).
%To get a bound on the term $|y_{k+2}+c_{k+1}+\theta|_{\infty}$ we need to find bounds on the terms $|1-y_{k+1}|_{\infty}^{-1}$ and $|1+y_{k+1}|_{\infty}^{-1}$. 
%Using the assumption and the first part of the lemma that we just proved, we get 
%\begin{equation}
%\label{arch1-y}
%\frac{1}{|1-y_{k+1}|_{\infty}}<\frac{5}{4}\epsilon_k^{(1-\delta)}\quad\mbox{and}\quad
%\frac{1}{|1+y_{k+1}|_{\infty}}<\frac{5}{4}\epsilon_k^{(1-\delta)}.
%\end{equation}

We have
$
|1\pm y_{k+1}|_{\infty}^{-1}<\frac{5}{4}\epsilon_k^{(1-\delta)}$,
%\frac{1}{|1+y_{k+1}|_{\infty}}<\frac{5}{4}\epsilon_k^{(1-\delta)}.
so
$|y_{k+2}+c_{k+1}+\theta|_{\infty}\leq |a_{k+1}+b_{k+1}+c_{k+1}|_{\infty}/(2|1-y_{k+1}|_{\infty})+|a_{k+1}-b_{k+1}+c_{k+1}|_{\infty}/(2|1+y_{k+1}|_{\infty})+|1-\theta y_k|_{\infty}<\frac{1}{8}\epsilon_k^{1-2\delta}+\frac{1}{8}\epsilon_k^{1-2\delta}+\epsilon_k
<%3\epsilon_k^\delta+3\epsilon_k^\delta+3\epsilon_k^\delta=
9\epsilon_k^\delta$.
This gives
$10\epsilon_k^\delta\leq |c_{k+1}+\theta\pm\theta|_{\infty}=|(y_{k+2}+c_{k+1}+\theta)\pm\theta(1\mp\theta y_{k+2})|_{\infty}
%&\leq& |y_{k+2}+c_{k+1}+\theta|_{\infty}+|1+\theta y_{k+2}|_{\infty},\nonumber\\
< 9\epsilon_k^\delta+|1\mp\theta y_{k+2}|_{\infty}$.
Hence, 
$
|y_{k+2}+c_{k+1}+\theta|_{\infty}\le9\epsilon_k^\delta<9|1\mp\theta y_{k+2}|_{\infty}.
$
Now 
$10\epsilon_{k+2}<10\epsilon_{k+2}^\delta\leq |c_{k+1}+\theta\pm\theta|_{\infty}= |(y_{k+2}+c_{k+1}+\theta)\pm\theta(1\mp\theta y_{k+2})|_{\infty}\leq 10 |1\mp\theta y_{k+2}|_{\infty}$.
%Note that in the above inequalities we have $10\epsilon_{k+2}^\delta<10\epsilon_{k+2}^\delta$, 
%By symmetry, had we considered the case $|y_{k-1}|_{\infty}\geq |1-\theta y_k|_{\infty}^{-(1-\delta)}$, then we would have obtained $|1\pm \theta y_{k-2}|_{\infty}\geq \epsilon_{k-2}$. Therefore, the proof is completed for this lemma.
\hfill $\mbox{ }\hspace{0cm}\square$
%\end{description}

%%%%%%%%%%%%%%%% end of lemma1%%%%%%%%%%%%%%%%%%%%%%%%
%We explored in the above lemma and its proof how far or close the iterates of equation (\ref{main eqn}) are to the singularities $\pm 1$ of the equation, given that $c_n\not\equiv 0$, $\pm 2$. Note that the terms used to describe the distance here (far, close) are with respect to the fixed absolute value under consideration and with respect to $\epsilon_n$.  Now we focus on the main result of this subsection which is a consequence of Theorem \ref{thm3.1.1} below. The main result of this subsection is to show that if $c_n$ is not identically 0 or $\pm 2$, then equation (\ref{main eqn}) is not a candidate for Diophantine integrability given that it has an admissible solution. We state and prove Theorem \ref{thm3.1.1} next where we use Lemma \ref{lemma3.1.1} in the proof. \\

%%%%%%%%%%%%%%%%%%%% theorem 1%%%%%%%%%%%%%%%%%%%%%

We are now ready to prove the following.

\begin{theorem} \textit{Let $(y_n)\subset \mathbb{Q}\setminus \{-1,1\}$ be an admissible solution of the equation (\ref{main eqn}),
%\begin{equation*}
%y_{n+1}+y_{n-1}=\frac{a_n+b_ny_n+c_ny_n^2}{1-y_n^2},
%\end{equation*}
where $a_n, b_n$ and $c_n$ are rational functions of $n$ with $c_n\neq 0$ or $\pm 2$ and the right hand side of (\ref{main eqn}) is of
degree 2 in $y_n$  for all sufficiently large $n$. 
Then there exists an integer $r_0$ such that for all $r\geq r_0$ and  $F<2$,  the summed logarithmic height 
%\begin{equation*}
$$
h_r(y_n)=\sum_{n=r_0}^r h(y_n)=\sum_{n=r_0}^r\sum_{p\leq\infty}\log^+|y_n|_p,
$$
%\end{equation*}
satisfies
$h_r(y_n)\geq F^rD$ for some $D>0$.}
\label{thm3.1.1}
\end{theorem}
%\end{thm}
%%%%%%%%%%%%%%%%%%%%%%%%%% Proof theorem1 %%%%%%%%%%%%%%%%%%
%\begin{description}

\noindent{\bf Proof :}
%The strategy in this subsection is to 
We will show that
there is a number $\tau<2$ such that for each absolute value $|.|_p$ ($\forall p\leq \infty$) on $\mathbb{Q}$ and for all $r\ge r_0$,
\begin{equation}
\label{Dio-ineq}
\sum_{n=r_0}^r
	\left(
		\log^+\frac1{|1-y_n|_p}+\log^+\frac1{|1+y_n|_p}
	\right)
\le \tau \sum_{n=r_0}^{r+1} \log^+|y_n|_p.
\end{equation}
We can then sum this inequality over all absolute values to show that the summed logarithmic height grows exponentially. 

Fix a prime $p\le\infty$ and an integer $r_0> K$ 
and define the four sets
%.  For all $n> K$ and for all $\epsilon_n$ given by equation (\ref{defeps}) with $\theta=1$ or $-1$, we choose and define the four sets
% Since we are concerned with the behaviour of the solution $y_n$ near the singularities $\pm 1$, we start by finding an upper bound for the expression $\sum_{k=r_0}^r (\log^+\frac{1}{|1-y_k|_p}+\log^+\frac{1}{|1+y_k|_p})$ for a fixed absolute value. Once we get this upper bound, we could sum over all the $p$-adic and Archimedean absolute values $\forall p\leq \infty$ for both sides of the inequality. Then in the left hand side of the inequality $h_r(y_n)$ appears and we attain our result given in the statement of the theorem when $r\rightarrow \infty$. 
% First we define four sets of points (for a fixed absolute value $|.|_p$) by
\begin{eqnarray*}
\label{sets}
%\begin{split}
A_r^\pm &=& \{n\,:\, r_0\le n \le r\mbox{  and  } |1\mp y_n|_p<\epsilon_n\},\\
B_r^\pm &=&\{n\,:\, r_0\le n \le r\mbox{  and  } \epsilon_n\le |1\mp y_n|_p<1\},
%\end{split}
\end{eqnarray*}
where  $\epsilon_n$ is given by equation (\ref{defeps}).
We now show that $A_r^+\cap A_r^-=\emptyset$.  For any $n\in A_r^+$ we have $|1-y_n|<\epsilon_n$.  
%In the non-Archimedean case
If $p<\infty$,
%\begin{equation*}
$$
\epsilon_n\leq \epsilon_n^{\delta}\leq |2|_p\leq \max\{|1-y_n|_p,|1+y_n|_p\}=|1+y_n|_p,
$$
%\end{equation*}
so $n\not\in A_r^-$.  The same conclusion holds in the Archimedean case since $\epsilon_n<1$ and so
$$
1<|2|_{\infty}-\epsilon_n\leq|1-y_n|_{\infty}+|1+y_n|_{\infty}-\epsilon_n<|1+y_n|_{\infty}.
$$

%%%%%%%%%%%%% The sets A_1 and A_3 (part 1)%%%%%%%%%%%%%%%

Lemma \ref{lemma3.1.1} shows that for each $n\in A_r^\pm$, we can define $\sigma_n^\pm=-1$ or 1 such that $|y_{n+\sigma_n^\pm}|_p\geq \frac{1}{|1\mp y_n|_p^{1-\delta}}$.  Lemma \ref{lemma3.1.1} also shows that
%Also define $\widehat{\sigma}_m=-1$ or 1, depending on the location of $m$ in the set $A_3(r)$, so that if $|1+y_m|_p<\epsilon_m$, then $|y_{m+\widehat{\sigma}_m}|_p\geq \frac{1}{|1+y_m|_p^{1-\delta}}$. Note that for each $n \in A_1(r)$ we have $n+\sigma_n \neq m+\widehat{\sigma}_m$ $\forall m\in A_3(r)$ since the distance between any small terms $|1-\theta y_n|_p<\epsilon_n$ (for $\theta=-1$ or 1) is more than 2 steps (Lemma \ref{lemma3.1.1}). This implies 
$\{n+\sigma_n^+|n\in A_r^+\}\cap\{n+\sigma_n^-|n \in A_r^-\}=\emptyset$
%Therefore, we will not have double counting of points when we sum the terms in the expression defined in (\ref{split}). 
 and that
 \begin{eqnarray}
\label{Apm}
 && \sum_{k\in A_r^+}\log^+\frac{1}{|1-y_k|_p}+\sum_{k\in A_r^-}\log^+\frac{1}{|1+y_k|_p}\nonumber\\ 
 &\leq& \frac{1}{1-\delta}\Big(\sum_{k\in A_r^+}\log^+|y_{k+\sigma_k^+}|_p
  +\sum_{k\in A_r^-}\log^+|y_{k+{\sigma}_k^-}|_p\Big)\nonumber \\
   &\leq&\frac{1}{1-\delta}\sum_{k=r_0-1}^{r+1}\log^+|y_k|_p.
\end{eqnarray}

% \begin{equation}
%\label{A_1}
%\sum_{k\in A_r^\pm} \log^+\frac{1}{|1-y_k|_p}%= \sum_{k\in A_1(r)}\log^+(|1-y_k|_p^{-(1-\delta)})^{\frac{1}{1-\delta}}
%= \frac{1}{1-\delta} \sum_{k\in A_r^\pm}\log^+|1-y_k|_p^{-(1-\delta)}
%\leq \frac{1}{1-\delta}\sum_{k\in A_r^\pm}\log^+|y_{k+\sigma_k}|_p.
%\end{equation}

Recalling the definition of $\epsilon_k$ in (\ref{defeps}), we have
\begin{eqnarray}
\label{Bpm}
&&\sum_{k\in B_r^\pm}\log^+\frac{1}{|1\pm y_k|_p}\leq\sum_{k\in B_r^\pm}\log^+\epsilon_k^{-1}%=\sum_{k\in A_2(r)}\log^+(\epsilon_k^{-\delta})^{\frac{1}{\delta}},
%\nonumber\\
%&=& \frac{1}{\delta}\sum_{k\in B_r^\pm}\log^+\Big(\kappa_p\max\{|2|_p^{-1},\cdots,|c_{k-1}\pm 2|_p^{-1}\}\Big)\nonumber\\
%&\leq&\frac{1}{\delta}\sum_{k=r_0}^r\log^+\kappa_p+\log^+|2|_p^{-1}+\cdots+\log^+|c_{k-1}\pm 2|_p^{-1})=: M_p.
%&=& 
=
\frac{1}{\delta}\sum_{k\in B_r^\pm}\log^+\Big(\kappa_p\max_{x\in X_k}\{|x|_p\}\Big)\nonumber\\
&\leq&
\frac{1}{\delta}\sum_{k=r_0}^r\left(\log^+\kappa_p+\sum_{x\in X_k}\log^+|x|_p\right)=: M_p.
\end{eqnarray}
%Similarly, we have
%\begin{equation}
%\label{A_4}
%\sum_{k\in A_4(r)}\log^+\frac{1}{|1+y_k|_p} \leq M.
%\end{equation}
So from the inequalities (\ref{Apm}) and (\ref{Bpm}) we see that %for a fixed absolute value $|.|_p$ ($p\leq \infty$), 
for all primes $p\le\infty$,
$$
%\sum_{k=r_0}^{r}\left(\log^+\frac{1}{|1-y_k|_p}+\log^+\frac{1}{|1+y_k|_p}\right)
\sum_{k=r_0}^r \left(\log^+\frac{1}{|1-y_k|_p}+\log^+\frac{1}{|1+y_k|_p}\right)
%&=&\sum_{k\in A_1(r)}\log^+\frac{1}{|1-y_k|_p}+\sum_{k\in A_2(r)}\log^+\frac{1}{|1-y_k|_p}\nonumber\\
%&+&\sum_{k\in A_3(r)}\log^+\frac{1}{|1+y_k|_p}+\sum_{k\in A_4(r)}\log^+\frac{1}{|1+y_k|_p}\nonumber\\
\leq\frac{1}{1-\delta}\sum_{k=r_0-1}^{r+1}\log^+|y_k|_p+2M_p.
$$
To get the height, we sum over all the primes ($p\leq \infty$) which yields
\begin{equation}
\label{heightinq}
\sum_{k=r_0}^{r}h\left(\frac{1}{1-y_k}\right)+\sum_{k=r_0}^r h\left(\frac{1}{1+y_k}\right)
\leq
\frac{1}{1-\delta}\sum_{k=r_0-1}^{r+1}h(y_k)+R_r,
\end{equation}
where
$$
R_r=
\frac 2\delta\sum_{k=r_0}^r\sum_{x\in X_k}h(x)+\frac{r-r_0}{2}\log 10=o\left(\sum_{k=r_0-1}^{r+1} h(y_k)\right),
$$
where the second equality follows from our admissibility condition (\ref{admissible}).
Furthermore, 
%from basic properties of heights 
we have
$
\left|h\left((1-\theta y_k)^{-1}\right)-h(y_k)\right|_{\infty}\leq \log2$,
 where $\theta=1$ or $-1$. 
 So we see that the summed logarithmic height satisfies
%\begin{equation*}
$
h_{r+1}(y_k)\geq 2(1-\delta)h_r(y_k)+o(h_{r+1}(y_k))
$
%\end{equation*}
%Any function $R_r=o(h_r(y_k))$ satisfies the following inequality for any fixed $\nu>0$, there exists $r_0$ such that $|R_r|_\infty<\nu\cdot |h_r(y_k)|_\infty$ holds for all $r>r_0$.
% Using this fact in (\ref{expgrowth2}) and applying this recurrence relation repeatedly, we get 
and hence for any $\nu>0$ there is a constant $D>0$ such that
\begin{equation}
\label{expg1}
h_r(y_k)\geq \left(\frac{2(1-\delta)}{1+\nu}\right)^rD.
\end{equation}
For sufficiently small $\delta,\nu$, $1<F=\frac{2(1-\delta)}{1+\nu} <2$, which proves the theorem.   \hfill$\square$\\

%\end{description}

%%%%%%%%%%%%%%%%%%%%%% subsection 4.2 %%%%%%%%%%%%%%%%%%%%%%%
%\subsection{\underline{Case 2: $c_n\equiv 0$ in equation (\ref{main eqn})}}
%%%%%%%%%%%%%%%%%%%%%% new part %%%%%%%%%%%%%%%%%%%%%%%%
%\section{Special values of $c_n$}

%\subsection{$c_n\equiv 0$}

Now we consider the case in which $c_n$ vanishes identically, i.e. 
\begin{equation}
\label{c=0case} 
y_{n+1}+y_{n-1}=\frac{a_n+b_ny_n}{1-y_n^2}.
\end{equation}
% where $a_n+b_n$ and $a_n-b_n$ are nonzero for all $n>K$. 
 Our strategy is again to prove an inequality of the form (\ref{Dio-ineq}) with $\tau<2$.
%%%%%%%%%%%%%%%%%%% Our Strategy %%%%%%%%%%%%%%%%%%%%%%%%%
%Before we proceed to apply our strategy we set the assumptions and definitions that are used throughout this subsection.
% Note that the functions $a_n\pm b_n$ are not equal to zero $\forall n$, otherwise the right hand side of $(\ref{c=0case})$ is reducible, contradicting our assumption.    
The integer $K$ is chosen such that for all $n>K$, $a_n+ b_n$ and $a_n- b_n$ are nonzero and each of the expressions $\pm a_n+b_n-2b_{n+1}$, $\pm a_n+b_n-2b_{n-1}$ and $a_{n}\pm b_{n}\pm (\pm a_{n-2}+b_{n-2}-2b_{n-1})$ is either identically zero or for all $n>K$ it is nonzero. 

In the following definition of $\epsilon_n$ we take the maximum over a set for which we ignore those elements that are undefined (or infinite) and take the maximum of all the remaining finite elements.
For sufficiently small $\delta>0$ and 
for all $n> K$ we define $\epsilon_n$ by
\begin{eqnarray}
%\begin{split}
%fl 
\epsilon_n^{-\delta}=\kappa_p\max\Bigl\{&&|2|_p^{-1},|1/2|_p\cdot |a_n\pm b_n|_p,|1/2|_p^{-1}\cdot |a_n\pm b_n|_p^{-1}, |a_{n+1}|_p,|b_{n+1}|_p,\nn\\
&&|a_{n-1}|_p,|b_{n-1}|_p,|1/2|_p\cdot |a_{n+2}\pm b_{n+2}|_p,|1/2|_p\cdot |a_{n-2}\pm b_{n-2}|_p,\nn\\
&&|\pm a_n+b_n-2b_{n+1}|_p,|\pm a_n+b_n-2b_{n-1}|_p,|a_n\pm b_n|_p^{-1},\nn\\
&&|\pm a_n+b_n-2b_{n+1}|_p^{-1},|\pm a_n+b_n-2b_{n-1}|_p^{-1},|a_n\pm b_n|_p,\nn\\
&&|1/2|_p^{-1}\cdot |a_{n}\mp b_{n}\mp (\pm a_{n-2}+b_{n-2}-2b_{n-1})|_p^{-1}\Bigr\},\label{eps2}
%\end{split}
\end{eqnarray}
where $\kappa_p=1$ if $p<\infty$ and  $\kappa_\infty=10$. It is evident from the definition that $\epsilon_n\leq 1$ when $p<\infty$ and $p\neq 2$, while $\epsilon_n<1$ when $p=\infty$ or $p=2$. 
%%%%%%%%%%%%%%%%%% Execute the strategy %%%%%%%%%%%%%%%%%%%%%%%%%%

We again define the sets $A_r^\pm$ and $B_r^\pm$ as in (\ref{sets}).
The points of $A_r^\pm$ will be called $\pm 1$ points (since $y_n$ is close to $\pm 1$ with respect to the absolute value).  As in the proof of Theorem \ref{thm3.1.1}, %(where in this part of the proof we did not use the assumption $c_n\not\equiv 0$, so it is still valid here) that if 
it can be shown that if
$|1-\theta y_n|_p<\epsilon_n$ for $\theta=1$ or $-1$, then $|1+\theta y_n|_p\ge \epsilon_n$.  Hence
$A_r^+\cap A_r^-=\emptyset$.  
%Using the same argument as in the proof of Theorem \ref{thm3.1.1}, we see that t
The admissibility of $y_n$ then implies that 
$$
\sum_{n=r_0}^r
	\left(
		\log^+\frac1{|1-y_n|_p}+\log^+\frac1{|1+y_n|_p}
	\right)
=\sum_{n\in A_r^+}\log^+\!\frac1{|1-y_n|_p}+\!\!\!\sum_{n\in A_r^-}\log^+\!\frac1{|1+y_n|_p}
+\Phi_r,$$
where
%$$
%\Phi_r=\sum_{n\in B_r^+}\log^+\frac1{|1-y_n|_p}+\sum_{n\in B_r^-}\log^+\frac1{|1+y_n|_p}.
%$$
$\sum_{p\le\infty}\Phi_r=o(h_{r+1}(y_n))$.

%In order to bound 
%$$
%\sum_{n\in A_r^+}\log^+\frac1{|1-y_n|_p}+\sum_{n\in A_r^-}\log^+\frac1{|1+y_n|_p}
%$$
%by a multiple of $\sum_{n=r_0}^{r+1}\log^+|y_n|_p$, w
We construct a number of disjoint subintervals
containing only 1 points, -1 points and points where $y_n$ is sufficiently large to make a significant contribution to the right hand side of  the inequality (\ref{Dio-ineq}).  %These subintervals are called oscillating sequences.

%\newtheorem{def8}{Definition}[section]
%\begin{def8}
\begin{definition}
\textit{
Suppose that $|1-\theta y_k|_p<\epsilon_k$, for some $k\in\mathbb{Z}$ and $\theta=1$ or $\theta=-1$.  Then the \underline{oscillating sequence} S containing $k$ is the longest interval in $\mathbb{Z}$ (possibly unbounded) satisfying the following conditions.
\begin{enumerate}
\item If $k+2l\in S$ then $|1-(-1)^l\theta y_{k+2l}|_p<\epsilon_{k+2l}$;
\item If $\{k+2l-1,k+2l\}\subseteq S$, then 
	$|y_{k+2l-1}|_p\ge |1-(-1)^l\theta y_{k+2l}|_p^{-(1-\delta)}$; and
\item If $\{k+2l,k+2l+1\}\subseteq S$, then 
	$|y_{k+2l+1}|_p\ge |1-(-1)^l\theta y_{k+2l}|_p^{-(1-\delta)}$.
\end{enumerate}
 }
 \end{definition}
%\end{def8}

%As in the proof of Lemma \ref{lemma3.1.1}, i
If  $|1-\theta y_n|_p<\epsilon_n$ then either $|y_{n+1}|_p\geq |1-\theta y_n|_p^{-(1-\delta)}$ or $|y_{n-1}|_p\geq |1-\theta y_n|_p^{-(1-\delta)}$, 
%(where for this part of the proof we did not use the assumption $c_n\not \equiv 0$, so it is still valid here). 
so every $\pm1$ point lies in an oscillating sequence containing at least two elements.  For a fixed oscillating sequence $S$ and $r\ge r_0$, we will now obtain a suitable upper bound for 
\begin{equation}
\label{sumS}
\sum_{n\in S\cap A_r^+}\log^+\frac1{|1-y_n|_p}+\sum_{n\in S\cap A_r^-}\log^+\frac1{|1+y_n|_p}.
\end{equation}

\vskip 5mm

\noindent{\bf Case 1:} Let $m+1$ be the total number of 1 points and -1 points in $S\cap[r_0,r]$ and assume that $m\ge 2$.
Let $I$ be the shortest subinterval of $S\cap[r_0,r]$ containing these $\pm 1$ points.  Let $k$ be the first term in $I$, so that $|1-\theta y_k|_p<\epsilon_k$ for some choice of $\theta=-1$ or 1.  Then $I=\{k,k+1,\ldots,k+2m\}$ and contains exactly $m$ points on which $y_n$ is big in the  sense that $|y_{k+1}|_p\ge |1-\theta y_k|_p^{-(1-\delta)}$,
$|y_{k+2m-1}|_p\ge |1-(-1)^m\theta y_{k+2m}|_p^{-(1-\delta)}$
and
$
|y_{k+2l+1}|_p
\ge
\max\{
	|1-(-1)^l\theta y_{k+2l}|_p^{-(1-\delta)},
	\allowbreak
	|1-(-1)^{l+1}\theta y_{k+2l+2}|_p^{-(1-\delta)}
	\}$,
for all $ l=1,\ldots,m-2$.
Hence
\begin{eqnarray*}
%\fl
 \sum_{n\in S\cap A_r^+}\log^+\frac1{|1-y_n|_p}+\sum_{n\in S\cap A_r^-}\log^+\frac1{|1+y_n|_p}
\\
=
\sum_{l=0}^m\log^+\frac{1}{|1-(-1)^l\theta y_{k+2l}|_p}
%\end{split}
%\end{equation*}
\\
%\begin{equation*}
%\begin{split}
=
\sum_{l=1}^m\frac lm \log^+\frac{1}{|1-(-1)^l\theta y_{k+2l}|_p}
+
\sum_{l=0}^{m-1}\frac {m-l}m \log^+\frac{1}{|1-(-1)^l\theta y_{k+2l}|_p}
\\
=
\sum_{l=0}^{m-1}\frac {l+1}m \log^+\frac{1}{|1-(-1)^{l+1}\theta y_{k+2l+2}|_p}
+
\sum_{l=0}^{m-1}\frac {m-l}m \log^+\frac{1}{|1-(-1)^l\theta y_{k+2l}|_p}
\\
\le
\frac 1{1-\delta}\sum_{l=0}^{m-1}\left(
							\frac{l+1}{m}+\frac{m-l}{m}
						\right)
\log^+|y_{k+2l+1}|_p
\\
=\frac{m+1}{(1-\delta)m}\sum_{l=0}^{m-1}
\log^+|y_{k+2l+1}|_p=
\frac{m+1}{(1-\delta)m}\sum_{n\in S\cap [r_0,r]}\log^+|y_n|_p
\\
\le
\frac{3}{2(1-\delta)}\sum_{n\in S\cap [r_0,r]}\log^+|y_n|_p,
%\end{split}
\end{eqnarray*}
where the last inequality follows from $m\ge 2$.

\vskip 5mm

\noindent{\bf Case 2:}
There are exactly two $\pm 1$ points in $S\cap[r_0,r]$.
Define $k$ such that these points are $k$ and $k+2$.  That is, for some choice of $\theta=\pm 1$, we have
$|1-\theta y_k|_p<\epsilon_k$ and $|1+\theta y_{k+2}|_p<\epsilon_{k+2}$.
We will use the following corollary of Theorem \ref{SC}. 
%{\bf We will prove Theorem \ref{SC} and this corollary at the end of this section.}

% \newtheorem{maincor}{Corollary}[section]
%\begin{maincor}
\begin{corollary}
\label{cor3.2.1}
\textit{ Fix a prime $p\leq \infty$.  For some $k>K$ let $\epsilon_k$ be given (\ref{eps2})  and 
suppose that for $\theta=1$ or $-1$, $|1-\theta y_k|_p<\epsilon_k$, $|y_{k-1}|_p\leq |1-\theta y_k|_p^{-1/2}$
and $|1+\theta y_{k+2}|_p<\epsilon_{k+2}$. Assume that
$a_{k}-\theta b_{k}-\theta (\theta a_{k-2}+b_{k-2}-2b_{k-1})\not\equiv 0$, then
$|y_{k+3}|_p>|1+\theta y_{k+2}|_p^{-1/2}$.}
\end{corollary}
%\end{maincor}

%%%%%%%%%%%%%%%%%%%%%%%%%%%%%%%% proof corollary 3.2.1 %%%%%%%%%%%%%%%%%%%%%%%%%%%%%%%%%%%%%%%%%%%%%%%%%%%%%%%

%\begin{description}

\noindent{\bf Proof:}
%In this proof, we use (\ref{eps2}) for $\epsilon_n^{-\delta}$. 
% Recall from Theorem \ref{SC} the following expression for $y_{k+3}$:
%\begin{equation}
%\label{expyk+3}
%y_{k+3}=\frac{\left(a_{k+2}-\theta b_{k+2}-\theta (\theta a_k+b_k-2b_{k+1})\right)}{2(1+\theta y_{k+2})}+C_k.
%\end{equation}
We begin with the non-Archimedean case $p<\infty$.  From part {\it (iii)} of Theorem \ref{SC} and the definition (\ref{eps2}) , we have
\begin{eqnarray}
\label{relation1}
 |1+\theta y_{k+2}|_p^{-(1-\delta)}
%=\frac{|1+\theta y_{k+2}|_p^{\delta}}{|1+\theta y_{k+2}|_p}
&<&\frac{\epsilon_{k+2}^{\delta}}{|1+\theta y_{k+2}|_p}
%\nonumber\\&\leq& 
\leq\frac{|a_{k+2}-\theta b_{k+2}-\theta (\theta a_k+b_k-2b_{k+1})|_p}{|2|_p\cdot |1+\theta y_{k+2}|_p
\nonumber
}
\\&=&
|y_{k+3}-C_k|_p\leq \max \{|y_{k+3}|_p, |C_k|_p\}.
\end{eqnarray}
From Theorem \ref{SC}  we have that for sufficiently small $\delta>0$, $|C_k|_p\leq |1+\theta y_{k+2}|_p^{-2/3-2\delta}\leq |1+\theta y_{k+2}|_p^{-(1-\delta)}$. So (\ref{relation1}) reduces to $|y_{k+3}|_p> |1+\theta y_{k+2}|_p^{-(1-\delta)}\geq |1+\theta y_{k+2}|_p^{-1/2}$. 

% The last part of the proof is f
 For the Archimedean absolute value, $\kappa_{\infty}=10$ in (\ref{eps2}), giving
 % where we had used (\ref{expyk+3}) and the triangle inequality: 
 \begin{eqnarray*}
%\label{relation2}
&&10|1+\theta y_{k+2}|_p^{-(1-\delta)}<\frac{10\epsilon_{k+2}^{\delta}}{|1+\theta y_{k+2}|_p} \\
&\leq& \frac{|a_{k+2}-\theta b_{k+2}-\theta (\theta a_k+b_k-2b_{k+1})|_p}{|2|_p\cdot |1+\theta y_{k+2}|_p}%\nonumber\\
=|y_{k+3}-C_k|_p\leq |y_{k+3}|_p+|C_k|_p\\
&\leq&|y_{k+3}|_p+2|1+\theta y_{k+2}|_p^{-2/3-2\delta}%\nonumber\\
\leq |y_{k+3}|_p+9|1+\theta y_{k+2}|_p^{-(1-\delta)}.
\end{eqnarray*}
So $|y_{k+3}|_p>|1+\theta y_{k+2}|_p^{-(1-\delta)}\geq |1+\theta y_{k+2}|_p^{-1/2}$ which completes the proof.
%proves the corollary for the Archimedean absolute value.
\hfill $\square$ \\  
%\end{description}
%%%%%%%%%%%%%%%%%%%%%%%%%%%%%%%%%%%%%%%%%% end of corollary 3.2.1 proof %%%%%%%%%%%%%%%%%%%%%%%%%%%%%%% 

Hence if $a_{k}-\theta b_{k}-\theta (\theta a_{k-2}+b_{k-2}-2b_{k-1})\not\equiv 0$, then either $|y_{k-1}|_p>|1-\theta y_k|_p^{-1/2}$ or
$|y_{k+3}|_p>|1+\theta y_{k+2}|_p^{-1/2}$. This says that, even if neither $k-1$ nor $k+3$ is in $S$, at least one of $y_{k-1}$ or $y_{k+3}$ has to be moderately large.  Without loss of generality, we assume that 
$|y_{k-1}|_p>|1-\theta y_k|_p^{-1/2}$.
For $\eta>0$,
we have
\begin{eqnarray}
\label{eta}
&& \sum_{n\in S\cap A_r^+}\log^+\frac1{|1-y_n|_p}+\sum_{n\in S\cap A_r^-}\log^+\frac1{|1+y_n|_p}
\nonumber
\\
&=&\log^+\frac{1}{|1-\theta y_k|_p}+\log^+\frac{1}{|1+\theta y_{k+2}|_p}\nonumber\\
&=&\eta \log^+\frac{1}{|1-\theta y_{k}|_p}
+(1-\eta)\log^+\frac{1}{|1-\theta y_k|_p}+\log^+\frac{1}{|1+\theta
y_{k+2}|_p}\nonumber\\
&\leq&
2\eta\log^+|y_{k-1}|_p
+\frac{1-\eta}{1-\delta}\log^+|y_{k+1}|_p+\frac{1}{1-\delta}\log^+|y_{k+1}|_p
\nonumber\\
&=&
2\eta\log^+|y_{k-1}|_p
+\frac{2-\eta}{1-\delta}\log^+|y_{k+1}|_p
.
\end{eqnarray}
So we can reduce the coefficient of $\log^+|y_{k+1}|_p$ by introducing a contribution from $y_{k-1}$.  If $k-1\in S$, this is not problematic and an upper bound for (\ref{eta}) is
$$
\max \left(
		\frac{2-\eta}{1-\delta},2\eta
	\right)\sum_{n\in S}\log^+|y_n|_p.
$$
However, if $k-1\not\in S$ then we need to be careful because we will later sum our estimates for 
(\ref{sumS}) over all oscillating sequences.  When we do this we might need to ``share'' the term $k-1$ with another oscillating sequence, in which case it will appear twice in the upper bound and we will need to sum the contributions.  Note that the term $k-1$ here cannot be part of a subinterval $I$ of the type considered in case 1 above as such subintervals of oscillating sequences have only $\pm 1$ points as endpoints.  There could, however, be two adjacent oscillating sequences $S_1$ and $S_2$ both of the type considered in the present case (case 2) that need to share the contribution from $y_{k-1}$.  If so, then summing over the contributions for both oscillating sequences would give the upper bound
$$
\frac{2-\eta}{1-\delta}\log^+|y_{k-3}|_p
+4\eta\log^+|y_{k-1}|_p
+\frac{2-\eta}{1-\delta}\log^+|y_{k+1}|_p
$$
which, in turn, is bounded from above by
$$
\max \left(
		\frac{2-\eta}{1-\delta},4\eta
	\right)\sum_{n=k-3}^{k+1}\log^+|y_n|_p.
$$
Note that $k-1$ could also be part of an oscillating sequence of the type we are about to consider in case 3.

\vskip 5mm

\noindent{\bf Case 3:}
There is exactly one $k_1\in S\cap[r_0,r]$ such that $|1-\theta y_{k_1}|_p <\epsilon_{k_1}$ for $\theta=-1$ or 1.  Since $S$ has at least two points, we know that either
$|y_{k_1-1}|_p\ge |1-\theta y_{k_1}|_p^{-(1-\delta)}$ or $|y_{k_1+1}|_p\ge |1-\theta y_{k_1}|_p^{-(1-\delta)}$.  Without loss of generality, we assume the latter.  Note that since $k_1\in S\cap[r_0,r]$,
then $k_1+1\in S\cap[r_0,r+1]$. 
So
$$
\sum_{n\in S\cap A_r^+}\!\!\!\log^+\frac1{|1-y_n|_p}+\!\!\!\!\sum_{n\in S\cap A_r^-}\!\!\!\log^+\frac1{|1+y_n|_p}
=\log^+\!\frac1{|1-\theta y_{k_1}|_p}
\le \frac 1{1-\delta}\log^+\!|y_{k_1+1}|_p.
$$
It is conceivable that $k_1+1$ is adjacent to, or part of, a sequence of the type considered in case 2 in such a way that it plays the role of $k-1$ in the analysis above of that case.  In other words, summing over the contributions of these two oscillating sequences in the left side of (\ref{Dio-ineq}) leads to a term of the form 
$$
\left(\frac{1}{1-\delta}+2\eta\right)\log^+|y_{k_1+1}|_p
$$
on the right hand side.

%We have now considered all possible oscillating sequences under the assumption that both
If both
$a_{k}-b_{k}- a_{k-2}-b_{k-2}+2b_{k-1}$ and $a_{k}+ b_{k}- a_{k-2}+b_{k-2}-2b_{k-1}$ are nonzero then combining our results
from the above cases, we have 
$$
\sum_{n=r_0}^r
	\left(
		\log^+\frac1{|1-y_n|_p}+\log^+\frac1{|1+y_n|_p}
	\right)
\le \tau \sum_{n=r_0}^{r+1} \log^+|y_n|_p+\Phi_r,
$$
where
$$
\tau=
\max
\left(
\frac{3}{2(1-\delta)},\frac{2-\eta}{1-\delta},2\eta,4\eta,\frac 1{1-\delta}+2\eta
\right).
$$
In particular, choosing $\eta=3/8$ and $\delta$ sufficiently small, we have
$\tau=3/4+(1-\delta)^{-1}<2$.
Since $ \sum_{p\le\infty}\Phi_r=o(h_{r+1}(y_n))$, we see that 
%%%%%%%%%%%%%%%%%%%%% end of the strategy %%%%%%%%%%%%%%%%%%%%%
\begin{eqnarray}
\label{grow1}
h_r(y_n)&\leq&\frac{\tau}{2}h_{r+1}(y_n)+o(h_{r+1}(y_n)),
\end{eqnarray}
so $h_r(y)$ grows exponentially with r.
     
%%%%%%%%%%%%%%%% special oscillating seciences %%%%%%%%%%%%%%%%%%%%
The argument above is based on the fact that Corollary \ref{cor3.2.1} guarantees that if $a_{k}-\theta b_{k}-\theta (\theta a_{k-2}+b_{k-2}-2b_{k-1})\not\equiv 0$ then there can be no {\em special oscillating sequences} as defined below.

%\newtheorem{def9}[def8]{Definition}
%\begin{def9}
\begin{definition}
\textit{The special oscillating sequence $S_p$ starting with $k$ is $S_p=\{k,k+1,k+2\}$. It is an oscillating sequence of length 3 starting with  $k$ in $\mathbb{Z}$ such that $|1-\theta y_k|_p<\epsilon_k$, $|y_{k+1}|_p\geq \max \Big\{|1-\theta y_k|_p^{-(1-\delta)}, |1+\theta y_{k+2}|_p^{-(1-\delta)}\Big\}$ and $|1+\theta y_{k+2}|_p<\epsilon_{k+2}$. Also, we have $|y_{k-1}|_p\leq |1-\theta y_k|_p^{-1/2}$ and $|y_{k+3}|_p\leq  |1+\theta y_{k+2}|_p^{-1/2}$. \\ }
\end{definition}
%\end{def9}

Note that there are two types of special oscillating sequences depending on whether $\theta=1$ or $\theta=-1$.
In order for $h_r(y_n)$ to grow sub-exponentially,  there must be infinitely many special oscillating sequences.  
If there are infinitely many special oscillating sequences of both types then both 
$a_{k}-\ b_{k}- ( a_{k-2}+b_{k-2}-2b_{k-1})$ and $a_{k}+ b_{k}+ (- a_{k-2}+b_{k-2}-2b_{k-1})$
must vanish, which characterises part {\it (i)} of the theorem.
The rest of this section will be a careful analysis of the case in which there are infinitely many 
special oscillating sequences of one kind only, corresponding to a fixed value of $\theta=\pm1$. 
For the rest of this section when we refer to special oscillating sequences we mean those sequences of the form $\theta$, $\infty$, $-\theta$,
for this fixed value of $\theta$ (where ``$\infty$'' refers to a large term). 

We define $f_n$ by
\begin{equation}
\label{ffunction}
f_n=(1-\theta y_n)y_{n+1}-y_n.
\end{equation}
So
$y_{n+1}=(f_n+y_n)/(1-\theta y_n)$, $y_{n-1}=(y_n-f_{n-1})/(1+\theta y_n)$, and
 (\ref{c=0case}) yield
$$
y_{n+1}+y_{n-1}=\frac{(f_n-f_{n-1})+(2+\theta f_n+\theta f_{n-1})y_n}{1-y_n^2}=\frac{a_n+b_ny_n}{1-y_n^2}.
$$%}
Hence
\begin{equation}
\label{ricc1}
(b_n-2-\theta f_n-\theta f_{n-1})y_n=f_n-f_{n-1}-a_n.
\end{equation}
If for all $n$, $b_n-2-\theta f_n-\theta f_{n-1}=0$,  then  $f_n-f_{n-1}-a_n=0$
and
%. Solving these two equations together to get $f_n$ we have
$$
f_n=\frac{1}{2\theta}(\theta a_n+b_n-2).
$$
%Since the expression of $f_n$ is in terms of the coefficients of equation (\ref{c=0case}), the summed logarithmic height $h_r(f_n)$ is a slow growing function with respect to $h_r(y_n)$ for an admissible solution $y_n$ of (\ref{c=0case}). This leads us to a discrete Riccati equation of the form
This shows that $y_n$ solves the discrete Riccati equation (\ref{ourriccati}).

Next consider the case $b_n-2-\theta f_n-\theta f_{n-1}\neq 0$, $\forall n>K$. From (\ref{ricc1}) we have
\begin{equation}
\label{ricc2}
y_n=\frac{f_n-f_{n-1}-a_n}{b_n-2-\theta f_n-\theta f_{n-1}}.
\end{equation}
%Note that we take $n$ to be greater than all the zeros of the rational function $b_n-2-\theta f_n-\theta f_{n-1}$. %which is valid since any non-zero rational function has a finite number of zeros, as stated earlier. 
Taking the logarithmic height of both sides of (\ref{ricc2}) and using some elementary properties of heights,
we have
\begin{eqnarray*}
h(y_n)&=&h\left(\frac{f_n-f_{n-1}-a_n}{b_n-2-\theta f_n-\theta f_{n-1}}\right)\\
&\leq&h(f_n-f_{n-1}-a_n)+h\left(\frac{1}{b_n-2-\theta f_n-\theta f_{n-1}}\right)\\
&=&h(f_n-f_{n-1}-a_n)+h(b_n-2-\theta f_n-\theta f_{n-1})\\
&\leq&2h(f_n)+2h(f_{n-1})+h(a_n)+h(b_n)+\log 24.
\end{eqnarray*}
Summing both sides of the inequality above and using the fact that $h_r(f_n)$ is a non-decreasing function of $n$, we have 
\begin{equation}
\label{ricinq}
h_r(y_n)\leq 4h_{r+1}(f_n)+h_r(a_n)+h_r(b_n)+(r-r_0+1)\log 24.
\end{equation}
%So $h_r(y_n)$ is bounded from above by $h_{r+1}(f_n)$ for an admissible solution $y_n$. Recall that for an admissible solution $y_n$ of (\ref{c=0case}), $h_r(a_n)$ and $h_r(b_n)$ are growing slower than $h_r(y_n)$. If $h_{r+1}(f_n)$ grows much slower than $h_{r+2}(y_n)$, then we show it leads to a fast growth of $h_{r}(y_n)$ with $r$. Before we proceed to investigate the relation between $h_{r+2}(y_n)$ and $h_{r+1}(f_n)$, we need to prove Lemma \ref{logest} that we use later in our argument.

%%%%%%%%%%%%%%%%%%%% end of lemma log measures %%%%%%%%%%%%%%

%Now we investigate the relation between $h_{r+1}(f_n)$ and $h_{r+2}(y_n)$.
From (\ref{ffunction}) we have
\begin{equation}
\label{fn}
f_n+\theta=\theta(1-\theta y_n)(1+\theta y_{n+1}).
\end{equation}
For every prime $p\leq \infty$, we define a set $C_p\subset \mathbb{Z}$ such that it consists of all the big terms in special oscillating sequences i.e. the terms $\infty$s in the form: $\theta$, $\infty$, $-\theta$. For a fixed prime $p$ and sufficiently large $r_0$, we have 
\begin{eqnarray}
%\begin{split}
%\fl 
\sum_{n=r_0}^r\log^+\frac{1}{|f_n+\theta|_p}=&\sum_{{\scriptsize \begin{array}{c} n=r_0\nn\\
 n\in C_p\end{array}}}^r\log^+\frac{1}{|f_n+\theta|_p}+\sum_{{\scriptsize \begin{array}{c} n=r_0\\
 n+1\in C_p\end{array}}}^r\log^+\frac{1}{|f_n+\theta|_p}
 \\
 &+\!\!\!\!\!\!\!\!\!\sum_{{\scriptsize \begin{array}{c} n=r_0\\
n\not\in C_p\mbox{ and }n+1\not\in C_p \end{array}}}^r\!\!\!\!\!\!\!\!\!\log^+\frac{1}{|f_n+\theta|_p}. \label{sumfn}
%\end{split}
\end{eqnarray}
In the above inequality we split the interval $[r_0,r]$ into points that are in special oscillating sequences (where $n$, $n+1\in C_p$) and points in any other oscillating sequence that is not special. Note that for 
$n\in C_p$, we have $|1+\theta y_{n+1}|_p^{-(1-\delta)}\leq |y_n|_p$. Therefore, for $n\in C_p$ we have
\begin{eqnarray*}
%\begin{split}
%\fl 
\log^+\frac{1}{|f_n+\theta|_p}&=&\log^+\frac{1}{|1-\theta y_n|_p\cdot |1+\theta y_{n+1}|_p}\leq \log^+|1-\theta y_n|_p^{-1}\cdot |y_n|_p^{\frac{1}{1-\delta}}\\
&=&\log^+|1-\theta y_n|_p^{-1}\cdot |y_n|_p^{\frac{\delta+1-\delta}{1-\delta}}=\log^+|1-\theta y_n|_p^{-1}\cdot|y_n|_p\cdot|y_n|_p^{\frac{\delta}{1-\delta}}\\
&\leq&\frac{\delta}{1-\delta}\log^+|y_n|_p+\log^+\left|\frac{y_n}{1-\theta y_n}\right|_p.
%\end{split}
\end{eqnarray*}
Since $|y_n|_p$ is big, it is away from $\theta$ and $-\theta$. If $p<\infty$, then $|y_n|_p=|\theta-\theta(1-\theta y_n)|_p\leq \max\{1,|1-\theta y_n|_p\}=|1-\theta y_n|_p$, since $|y_n|_p>1$. Hence, the term $\log^+\left|\frac{y_n}{1-\theta y_n}\right|_p$ vanishes. For $p=\infty$ we have the following relation $\epsilon_{n+1}^{-\delta}<\epsilon_{n+1}^{-(1-\delta)}<|1+\theta y_{n+1}|_{\infty}^{-(1-\delta)}\leq |y_n|_{\infty}\leq 1+|1-\theta y_n|_{\infty}$ which yields $\epsilon_{n+1}^{-\delta}-1\leq |1-\theta y_n|_{\infty}$. Consequently, $\frac{1}{|1-\theta y_n|_{\infty}}\leq \frac{1}{\epsilon_{n+1}^{-\delta}-1}$. Starting with $|y_n|_{\infty}\leq 1+|1-\theta y_n|_{\infty}$ then dividing both sides by $|1-\theta y_n|_{\infty}$ implies $\frac{|y_n|_{\infty}}{|1-\theta y_n|_{\infty}}\leq \frac{1}{|1-\theta y_n|_{\infty}}+1\leq \frac{1}{\epsilon_{n+1}^{-\delta}-1}+1$. Therefore, $\left|\frac{y_n}{1-\theta y_n}\right|_{\infty}\leq \frac{1}{4}+1=\frac{5}{4}$ since $5\leq \epsilon_{n+1}^{-\delta}$, giving
\begin{equation}
\label{ncp}
\sum_{p\leq \infty}\sum_{{\scriptsize \begin{array}{c} n=r_0\\
 n\in C_p\end{array}}}^r\log^+\frac{1}{|f_n+\theta|_p}\leq \frac{\delta}{1-\delta}h_{r}(y_n)+(r-r_0+1)\log (5/4).
\end{equation}
%where $V=\log\frac{5}{4}$. 
Similarly,
\begin{equation}
%\fl 
\label{n+1cp}
\sum_{p\leq \infty}\sum_{{\scriptsize \begin{array}{c} n=r_0\\
 n+1\in C_p\end{array}}}^r\log^+\frac{1}{|f_n+\theta|_p}\leq \frac{\delta}{1-\delta}h_{r+1}(y_n)+(r-r_0+1)\log (5/4).
\end{equation}
Summing over all $p\leq \infty$ in (\ref{sumfn}) and using (\ref{fn}), (\ref{ncp}) and (\ref{n+1cp}) yields
\begin{eqnarray*}
%\begin{split}
%\fl 
h_r(f_n)-(r-r_0+1)\log 2\leq h_r\left(\frac{1}{f_n+\theta}\right)\leq \frac{2\delta}{1-\delta}h_{r+1}(y_n)+2(r-r_0+1)\log (5/4)\\
\ \ \ +\sum_{p\leq \infty}\left\{\sum_{{\scriptsize \begin{array}{c} n=r_0\\
 n+1\not\in C_p\end{array}}}^r\log^+\frac{1}{|1-\theta y_n|_p}+\sum_{{\scriptsize \begin{array}{c} n=r_0\\
 n\not\in C_p\end{array}}}^r\log^+\frac{1}{|1+\theta y_{n+1}|_p}\right\}.
%  \end{split}
\end{eqnarray*}
Therefore,
\begin{equation}
\label{fnyn}
 h_r(f_n)\leq\frac{2\delta}{1-\delta}h_{r+1}(y_n)+(r-r_0+1)\log (25/8)+B_{r+1}, 
\end{equation}
where
\begin{equation}
%\fl 
\label{bn}
{\ \ \ \ \hskip 1 cm }B_r=\sum_{p\leq \infty}\left\{\sum_{{\scriptsize \begin{array}{c} n=r_0\\
 n+1\not\in C_p\end{array}}}^r\log^+\frac{1}{|1-\theta y_n|_p}+\sum_{{\scriptsize \begin{array}{c} n=r_0\\
 n-1\not\in C_p\end{array}}}^r\log^+\frac{1}{|1+\theta y_{n}|_p}\right\}.
\end{equation}

From our previous analysis of oscillating sequences that are not special, it follows from (\ref{grow1}) that
\begin{equation}
\label{bnestimates}
B_r\leq\tau\sum_{p\leq \infty}\sum_{{\scriptsize \begin{array}{c} n=r_0\\
 n\not\in C_p\end{array}}}^{r+1}\log^+|y_n|_p+R_r.
\end{equation}
Recall that $\tau<2$ and $R_r$ is an expression that involves the summed logarithmic heights of the coefficients $a_n$ and $b_n$. Applying the shift $r\rightarrow r+1$ in (\ref{fnyn}) and (\ref{bnestimates}), then using the result in (\ref{ricinq}) yields
\begin{equation}
\label{eqn9}
h_r(y_n)\leq\frac{8\delta}{1-\delta}h_{r+2}(y_n)+4\tau\sum_{p\leq\infty}\sum_{{\scriptsize \begin{array}{c} n=r_0\\
 n\not\in C_p\end{array}}}^{r+3}\log^+|y_n|_p+\widehat{R}_{r+2},
\end{equation}
 where $\widehat{R}_{r+2}=o(h_{r+2}(y_n))$. 
 
 %If we could compare the expression $\sum_{p\leq\infty}\sum_{{\scriptsize \begin{array}{c} n=r_0\\ n\not\in C_p\end{array}}}^{r+1}\log^+|y_n|_p$ to $h_{r+1}(y_n)$ such that it is smaller than a suitable constant $c$ times $h_{r+1}(y_n)$ in a big set of positive integers, then we show that $h_r(y_n)$ grows very fast with $r$. We construct another inequality that has $B_r$ and consequently, the expression $\sum_{p\leq \infty}\sum_{{\scriptsize \begin{array}{c} n=r_0\\ n\not\in C_p\end{array}}}^{r+1}\log^+|y_n|_p$. If in this constructed inequality, $\sum_{p\leq \infty}\sum_{{\scriptsize \begin{array}{c} n=r_0\\ n\not\in C_p\end{array}}}^{r+1}\log^+|y_n|_p$ is bounded from below by $c$ times $h_{r+1}(y_n)$ in a big set of positive integers, then we show that this also leads to a very fast growth of $h_r(y_n)$ with $r$. %We use Lemma \ref{logest} to show the above result.
 
 Now we consider the following inequality
 \begin{eqnarray*}
 %\begin{split}
&&\sum_{p\leq\infty}\sum_{n=r_0}^r\left\{\log^+\frac{1}{|1-y_n|_p}+\log^+\frac{1}{|1+y_n|_p}\right\}\\
&\leq&\sum_{p\leq\infty}\left\{\sum_{{\scriptsize \begin{array}{c} n=r_0\\
 n+1\in C_p\end{array}}}^r\log^+\frac{1}{|1-\theta y_n|_p}+\sum_{{\scriptsize \begin{array}{c} n=r_0\\
 n-1\in C_p\end{array}}}^r\log^+\frac{1}{|1+\theta y_n|_p}\right\}+B_r.
%\end{split}
\end{eqnarray*}
Recall that if $n+1\in C_p$ (or $n-1\in C_p$), then $|y_{n+1}|_p\geq |1-\theta y_n|_p^{-(1-\delta)}$ (or $|y_{n-1}|_p\geq |1+\theta y_n|_p^{-(1-\delta)}$). Using this fact and (\ref{bnestimates}) we have 
\begin{eqnarray}
\label{eqn10}
&&\sum_{p\leq\infty}\sum_{n=r_0}^r\left\{\log^+\frac{1}{|1-y_n|_p}+\log^+\frac{1}{|1+y_n|_p}\right\}\nonumber\\
&\leq&\frac{2}{1-\delta}\sum_{p\leq\infty}\sum_{{\scriptsize \begin{array}{c} n=r_0\\
 n\in C_p\end{array}}}^{r+1}\log^+|y_n|_p+\tau\sum_{p\leq\infty}\sum_{{\scriptsize \begin{array}{c} n=r_0\\n\not\in C_p\end{array}}}^{r+1}\log^+|y_n|_p+R_r\nonumber\\
 &=&\frac{2}{1-\delta}\sum_{p\leq\infty}\sum_{{\scriptsize \begin{array}{c} n=r_0\\
 n\in C_p\end{array}}}^{r+1}\log^+|y_n|_p+\frac{2}{1-\delta}\sum_{p\leq\infty}\sum_{{\scriptsize \begin{array}{c} n=r_0\\
 n\not\in C_p\end{array}}}^{r+1}\log^+|y_n|_p\nonumber\\
 &-&\frac{2}{1-\delta}\sum_{p\leq\infty}\sum_{{\scriptsize \begin{array}{c} n=r_0\\
 n\not\in C_p\end{array}}}^{r+1}\log^+|y_n|_p+\tau\sum_{p\leq\infty}\sum_{{\scriptsize \begin{array}{c} n=r_0\\
 n\not\in C_p\end{array}}}^{r+1}\log^+|y_n|_p+R_r\nonumber\\
 &=&\frac{2}{1-\delta}h_{r+1}(y_n)-\left(\frac{2}{1-\delta}-\tau\right)\sum_{p\leq\infty}\sum_{{\scriptsize \begin{array}{c} n=r_0\\
 n\not\in C_p\end{array}}}^{r+1}\log^+|y_n|_p+R_r.
\end{eqnarray}
This implies that 
\begin{equation}
\label{eqn11}
2h_r(y_n)\leq \frac{2}{1-\delta}h_{r+1}(y_n)-\left(\frac{2}{1-\delta}-\tau\right)\sum_{p\leq\infty}\sum_{{\scriptsize \begin{array}{c} n=r_0\\
 n\not\in C_p\end{array}}}^{r+1}\log^+|y_n|_p+\widetilde{R_r},
\end{equation}
where $\widetilde{R_r}=o(h_{r}(y_n))$ as $r\to\infty$. 

Considering the two inequalities in (\ref{eqn9}) and (\ref{eqn11}), we have two cases to consider depending on whether the expression $\sum_{p\leq\infty}\sum_{{\scriptsize \begin{array}{c} n=r_0\\
 n\not\in C_p\end{array}}}^{r+1}\log^+|y_n|_p$ is very small compared to $h_{r+1}(y_n)$ on a large set. 
 In either case we obtain an inequality of the form $h_{r+s}(y_n)\ge \alpha h_r(y_n)$, for some $\alpha<1$ and $s>0$, on a set of infinite logarithmic measure, which implies conclusion {\it (iii)} of the theorem.
 %We use Lemma \ref{logest} to prove that in both cases $h_r(y_n)$ grows very fast with $r$. 
% \begin{itemize}
\\
\textbf{Case 1}: Assume that there is a sufficiently small constant $c>0$ such that 
$$
\sum_{p\leq\infty}\sum_{{\scriptsize \begin{array}{c} n=r_0\\
 n\not\in C_p\end{array}}}^{r+1}\log^+|y_n|_p\leq c h_{r+1}(y_n),
$$
on a set of infinite discrete logarithmic measure. Then (\ref{eqn9}) implies 
$$
h_r(y_n)\leq \left(\frac{8\delta}{1-\delta}+4\tau c\right)h_{r+3}(y_n)+\widehat{R}_{r+2},
$$
on a set of infinite discrete logarithmic measure. %Hence, Lemma \ref{logest} implies that $h_r(y_n)$ grows very fast with $r$.
\\
\textbf{Case 2}: Assume that 
$$
\sum_{p\leq \infty}\sum_{{\scriptsize \begin{array}{c} n=r_0\\
 n\not\in C_p\end{array}}}^{r+1}\log^+|y_n|_p> c h_{r+1}(y_n),
$$
on a set of infinite discrete logarithmic measure. Using this inequality in (\ref{eqn11}) yields
$$
2h_r(y_n)\leq \left[\frac{2}{1-\delta}-\left(\frac{2}{1-\delta}-\tau\right)c\right]h_{r+1}(y_n)+\widetilde{R}_r.
$$
Since $\left[\frac{2}{1-\delta}-(\frac{2}{1-\delta}-\tau)c\right]<2$ for sufficiently small $\delta$.
%, then using Lemma \ref{logest} the above inequality implies that $h_r(y_n)$ grows very fast with $r$.\\
%\end{itemize}
%This completes the proof of {\bf CHECK no Theorem 1.4.1} in the case $c_n\equiv 0$. 

% We use this lemma to show the fast growth of $h_r(y_n)$ with $r$ later. Clearly if a non-decreasing sequence of positive numbers $(w_n)$ satisfies $w_{n+s}\geq \alpha w_n$ $\forall n$, where $s>0$ and  $\alpha>1$, then $w_n$ grows exponentially. Lemma \ref{logest} says that if $(w_n)$ satisfies the inequality on a sufficiently large set (which has infinite discrete logarithmic measure), then $w_n$ still grows very fast. 

Conclusion {\it (iii)} of the theorem follows from the following %Lemma 
with $w_r=h_r(y_n)$.
%%%%%%%%%%%%%%%%%%%%%% lemma log measures %%%%%%%%%%%%%%%%%% 
%\newtheorem{logmeasure}[thm]{Lemma}
\begin{lemma}
%\begin{logmeasure}
\textit{Let $(w_n)_{n\geq n_0}$ $(n_0>0)$ be a non-decreasing sequence of positive numbers. For a fixed real number $\alpha>1$ and a fixed positive integer $s$ we define 
\begin{equation}
\label{settt}
F=\{n\geq n_0 : \mbox{ } w_{n+s}\geq \alpha w_n\}.
\end{equation}
If $F$ has infinite discrete logarithmic measure, i.e.
$\displaystyle
\sum_{n\in F}\frac{1}{n}=\infty$,
 then
 \begin{equation}
 \label{dislogmeasure}
\limsup_{r\to \infty} \frac{\log \log w_r}{\log r}\geq 1.
\end{equation}
}\label{logest}
\end{lemma}
%\end{logmeasure}
%\begin{description}

\noindent{\bf Proof:}
Define a sequence $(r_n)$ using induction as follows. Let $r_0=\min (F)$ and for all $n>0$, define $r_n=\min (F\cap [r_{n-1}+s,\infty))$. Hence, $r_{n+1}\geq r_n+s$ and 
$F\subseteq \cup_{n=0}^\infty [r_n,r_n+s]$.
This yields $w_{r_{n+1}}\geq w_{r_n+s}\geq \alpha w_{r_n}$ for all $n\geq 0$. Iterating this relation recursively yields
\begin{equation}
\label{westimate}
w_{r_{n}}\geq \alpha^n w_{r_0}.
\end{equation}
 We use the notation $\lfloor x\rfloor$ to denote the integer part of $x$ in the following chain of inequalities. Assume that there is a constant $\varepsilon >0$ and an integer $m>1$ such that $r_n\geq n^{1+\varepsilon}$ for all $n>m$. Then there is a constant $E$ such that 
 \begin{eqnarray*}
\sum_{j\in F}\frac{1}{j}&\leq& E+\sum_{n=m}^{\infty}\sum_{k=\lfloor n^{1+\varepsilon}\rfloor}^{\lfloor n^{1+\varepsilon}\rfloor+s}\frac{1}{k}
%\leq E+\sum_{n=m}^{\infty}\int_{\lfloor n^{1+\varepsilon}\rfloor-1}^{\lfloor n^{1+\varepsilon}\rfloor+s}\frac{\mbox{d}t}{t} \\
\leq E+\sum_{n=m}^{\infty}\int_{ n^{1+\varepsilon}-2}^{n^{1+\varepsilon}+s}\frac{\mbox{d}t}{t}\\
%&=&E+\sum_{n=m}^{\infty}\log t\Big|_{n^{1+\varepsilon}-2}^{n^{1+\varepsilon}+s}\\
%&=&E+\sum_{n=m}^{\infty}\log \left(\frac{n^{1+\varepsilon}+s}{n^{1+\varepsilon}-2}\right)\\
&\leq&E+\sum_{n=m}^{\infty}\left((s+2)n^{-(1+\varepsilon)}+O(n^{-2(1+\varepsilon)})\right)<\infty.
\end{eqnarray*}
But this is a contradiction to our assumption that $F$ has infinite discrete logarithmic measure. Therefore, there exists a subsequence $(r_{n_k})$ such that $r_{n_k}<n_k^{1+\varepsilon}$ for all $k\geq 0$. From (\ref{westimate}) we have 
$w_{r_{n_k}}\geq \alpha^{n_k}w_{r_0}$.
Hence, 
\begin{eqnarray*}
\limsup_{r\longrightarrow \infty}\frac{\log\log w_r}{\log r}&\geq&\limsup_{k\longrightarrow \infty}\frac{\log\log w_{r_{n_k}}}{\log r_{n_k}}
\geq\limsup_{k\longrightarrow \infty}\frac{\log\log \alpha^{n_k}w_{r_0}}{\log n_k^{1+\varepsilon}}\\
&=&\limsup_{k\longrightarrow \infty}\frac{\log\left(n_k\log\alpha+\log w_{r_0}\right)}{(1+\varepsilon)\log n_k}
\geq \frac{1}{1+\varepsilon}.
\end{eqnarray*}
Since $\varepsilon>0$ is arbitrary small number, this proves the lemma. \hspace{2.8cm}$\square$
%\end{description}
\vskip 4mm
This concludes the case in which $c_n\equiv 0$.  The cases $c_n\equiv\pm 2$ are similar except that the existence of Riccati solutions is prohibited by the assumption in the theorem that the degree of the right side of equation (\ref{main eqn})  is two.
%We end this subsection by giving the proofs of Theorem \ref{SC} and Corollary \ref{cor3.2.1}.\newline
%%%%%%%%%%%%%%%%%%%%%%% Proof of theorem 3.2.1 %%%%%%%%%%%%%%%%
% not included yet
%%%%%%%%%%%%%%%%%%%%%% subsection 4.3 %%%%%%%%%%%%%%%%%%%%%%%
%\subsection{\underline{$c_n\equiv 2$ or $c_n\equiv -2$ in equation (\ref{main eqn})}} 

\section{Proof of Theorem \ref{SC}}

For a fixed absolute value $|.|_p$, assume that $|1-\theta y_k|_p<\epsilon_k$ for some $k>K$, where $\theta=-1$ or 1.
First we rewrite equation (\ref{c=0case}) as
\begin{equation}
\label{eqsimple}
y_{k+1}+y_{k-1}=\frac{1/2(a_k+\theta b_k)}{1-\theta y_k}+\frac{1/2(a_k-\theta b_k)}{1+\theta y_k}.
\end{equation}
It follows that
\begin{equation}
\label{Akforms}
A_k:=y_{k+1}-\frac{\textstyle 1/2(a_k+\theta b_k)}{\textstyle 1-\theta y_k}=\frac{\textstyle 1/2(a_k-\theta b_k)}{\textstyle 1+\theta y_k}-y_{k-1}.
\end{equation}
We begin by considering non-Archimedean absolute values ($p<\infty$). In this case
 \begin{equation}
\label{Aestimate}
|A_k|_p\leq \max\left\{\frac{|1/2|_p\cdot |a_k-\theta b_k|_p}{|1+\theta y_k|_p},|y_{k-1}|_p\right\}.
\end{equation}
%Here we need to find an estimate for the terms in the set above to prove the estimate for $|A_k|_p$ in the theorem. For $p<\infty$, we have
Note that $
\epsilon_k^{\delta}\leq |2|_p\leq \max \{|1-\theta y_k|_p,|1+\theta y_k|_p\}$. 
 If  $|1-\theta y_k|_p>|1+\theta y_k|_p$, then we obtain the contradiction $\epsilon_k^{\delta}<\epsilon$.  Hence
$\epsilon_k^{\delta}\leq\max\{|1-\theta y_k|_p,|1+\theta y_k|_p\}=|1+\theta y_k|_p$.
This implies ${\textstyle |1+\theta y_k|_p}^{-1}\leq \epsilon_k^{-\delta}<|1-\theta y_k|_p^{-\delta}$. Using this relation and  (\ref{eps2}) in (\ref{Aestimate}) yields
$|A_k|_p%\leq \max\left\{\frac{|1/2|_p\cdot |a_k-\theta b_n|_p}{|1+\theta y_k|_p},|y_{k-1}|_p\right\}
\leq\max\left\{|1-\theta y_k|_p^{-2\delta},|1-\theta y_k|_p^{-1/2}\right\}=|1-\theta y_k|_p^{-1/2}$.
%The above result is valid for sufficiently small $\delta$. This proves the first part of the theorem for the non-Archimedean absolute value ($p<\infty$).\\

%To prove the second part, we start by writing the expression for $B_k$ as follows:
Consider
\begin{equation}
\label{Bk-direct}
B_k:=(y_{k+2}+\theta)-\left(\theta-\frac{2b_{k+1}}{a_k+\theta b_k}\right)(1-\theta  y_k).
%=\frac{a_{k+1}+b_{k+1}y_{k+1}}{(1-y_{k+1})(1+y_{k+1})}+\frac{b_{k+1}}{y_{k+1}-A_k}.
\end{equation}
Incrementing equation (\ref{c=0case}) we obtain
 \begin{eqnarray}
%\begin{split}
B_k= \frac{a_{k+1}}{(1-y_{k+1})(1+y_{k+1})}+\frac{b_{k+1}}{(1-y_{k+1})(1+y_{k+1})(y_{k+1}-A_k)}\nn\\
-\frac{b_{k+1}A_ky_{k+1}}{(1-y_{k+1})(1+y_{k+1})(y_{k+1}-A_k)}.\label{Bcase}
%\end{split}
\end{eqnarray}
%In order to get an estimate for $|B_k|_p$, we need an estimate of the three terms on the right hand side of equation (\ref{Bcase}). We start first with 
Now
\begin{equation}
%|1-\theta y_k|_p^{-(1-\delta)}=
{|1-\theta y_k|_p}^{-(1-\delta)}<\frac{\epsilon_k^{\delta}}{|1-\theta y_k|_p}\le  \frac{|1/2|_p\cdot |a_k+\theta b_k|_p}{|1-\theta y_k|_p}
=|y_{k+1}-A_k|_p.
\end{equation}
%{\bf Should be inequality above - check direction.} 
So
$|1-\theta y_k|_p^{-(1-\delta)}<
 \max\{|y_{k+1}|_p,|1-\theta y_k|_p^{-1/2}\}\leq \max\{|y_{k+1}|_p,|1-\theta y_k|_p^{-(1-\delta)}\}$ $=|y_{k+1}|_p$.
Hence,
$\epsilon_k^{-(1-\delta)}<|1-\theta y_k|_p^{-(1-\delta)}\leq |y_{k+1}|_p=|1-(1\pm y_{k+1})|_p\leq \max\{1,|1\pm y_{k+1}|_p\}
\le\max\{\epsilon^{-(1-\delta)},|1\pm y_{k+1}|_p\}$, giving
%Therefore, since $\epsilon_k\leq 1$, the maximum  is $|1-y_{k+1}|_p$. This implies 
\begin{equation}
\label{yk+1esta}
{|1\pm y_{k+1}|_p^{-1}}\leq |1-\theta y_k|_p^{1-\delta}.
\end{equation}
Moreover, we have from the first part of the theorem that
\begin{equation}
\label{yupper}
|y_{k+1}|_p\leq \max\left\{\frac{|1/2|_p\cdot |a_k+\theta b_k|_p}{|1-\theta y_k|_p},|A_k|_p\right\}
\leq |1-\theta y_k|_p^{-(1+\delta)}.
\end{equation}
%If we apply the non-Archimedean absolute value ($p<\infty$) to the 
Taking the $p$-adic absolute value of equation (\ref{Bcase}) and using the estimates above, we get
$$
|B_k|_p
%&\leq& 
%\max\Biggl\{\frac{|a_{k+1}|_p}{|1-y_{k+1}|_p\cdot |1+y_{k+1}|_p}, \frac{|b_{k+1}|_p}{|1-y_{k+1}|_p\cdot |1+y_{k+1}|_p\cdot |y_{k+1}-A_k|_p},\\
%&&\frac{|b_{k+1}|_p\cdot |A_k|_p\cdot |y_{k+1}|_p}{|1-y_{k+1}|_p\cdot |1+y_{k+1}|_p\cdot |y_{k+1}-A_k|_p}\Biggr\}\\
\leq\max\left\{|1-\theta y_k|_p^{2-3\delta},|1-\theta y_k|_p^{3-4\delta},|1-\theta y_k|_p^{3/2-5\delta}\right\}.
$$
Hence $|B_k|_p\leq |1-\theta y_k|_p^{3/2-5\delta}$, as required.
%\end{eqnarray*}
%Here we used the results in (\ref{yk+1esta}), (\ref{yk+1estb}), (\ref{y-A}) and (\ref{yupper}). Also, we used the result from the first part of the theorem and $\epsilon_k^{-\delta}$ definition in (\ref{eps2}). Hence, we proved the second part of the theorem for the non-Archimedean absolute value ($ p<\infty$).\\ 

Next we have
$$
C_k:=y_{k+3}-\frac{(a_{k+2}-\theta b_{k+2}-\theta (\theta a_k+b_k-2b_{k+1}))}{2(1+\theta y_{k+2})}.
$$
Incrementing equation (\ref{c=0case}) twice and eliminating $y_{k+3}$ from the above yields
$$
C_k=\frac{1/2(a_{k+2}+\theta b_{k+2})}{1-\theta y_{k+2}}-\frac{a_k+\theta b_k}{2(1-\theta y_k)}+\frac{\theta (\theta a_k+b_k-2b_{k+1})}{2(1+\theta y_{k+2})}-A_k.
$$
 Combining the two middle terms and using part (ii) %of the theorem %to eliminate $y_{k+2}$ 
 in the numerator gives
\begin{equation}
\label{Ckcase}
C_k=\frac{1/2(a_{k+2}+\theta b_{k+2})}{1-\theta y_{k+2}}-\frac{B_k(a_k+\theta b_k)}{2\theta (1+\theta y_{k+2})(1-\theta y_k)}-A_k.
\end{equation}
% In order to find an upper bound for $|C_k|_p$, we need to find upper bounds for each term in the right hand side of the equation in (\ref{Ckcase}). We start with the first term $\frac{\textstyle 1/2(a_{k+2}+\theta b_{k+2})}{\textstyle 1-\theta y_{k+2}}$. We need to find an upper bound for $\frac{\textstyle 1}{\textstyle |1-\theta y_{k+2}|_p}$. 
From part (ii) of the theorem, we have
 \begin{eqnarray}
 \label{k+2relk}
|1+\theta y_{k+2}|_p&\leq& \max\left\{\frac{|\theta a_k+b_k-2b_{k+1}|_p}{|a_k+\theta b_k|_p}|1-\theta y_k|_p, |B_k|_p\right\}\nonumber\\
&\leq&\max\{|1-\theta y_k|_p^{1-2\delta}, |1-\theta y_k|_p^{3/2-5\delta}\}
%\nonumber\\&<&%\max\{\epsilon_k^{1-2\delta},\epsilon_k^{3/2-5\delta}\}=
<\epsilon_k^{1-2\delta},
\end{eqnarray}
%In the above chain of inequalities for sufficiently small $\delta$, 
where we have used (\ref{eps2}).
% and the assumption $|1-\theta y_k|_p<\epsilon_k$. 
%Note that if $|\theta a_k+b_k-2b_{k+1}|_p\equiv 0$, then $|1+\theta y_{k+2}|_p=|B_k|_p\leq |1-\theta y_k|_p^{3/2-5\delta}<\epsilon_k^{3/2-5\delta}\leq \epsilon_k^{1-2\delta}$. 
Also, %we have for $p<\infty$ the following relation:
$\epsilon_k^{\delta}\leq |2|_p\leq \max\{|1+\theta y_{k+2}|_p,|1-\theta y_{k+2}|_p\}\le \max\{\epsilon_k^{1-2\delta},|1-\theta y_{k+2}|_p\}
=|1-\theta y_{k+2}|_p$.
%If we have $|1+\theta y_{k+2}|_p\geq|1-\theta y_{k+2}|_p$, then $\epsilon_k^{\delta}\leq \max\{|1+\theta y_{k+2}|_p,|1-\theta y_{k+2}|_p\}=|1+\theta y_{k+2}|_p<\epsilon_k^{1-2\delta}$. This is a contradiction, since $\epsilon_k^{1-2\delta}\leq \epsilon_k^{\delta}$ for sufficiently small $\delta$. Therefore, $|1-\theta y_{k+2}|_p> |1+\theta y_{k+2}|_p$. 
Hence
\begin{equation}
\label{1-yk+2}
{|1-\theta y_{k+2}|_p}^{-1}\leq \epsilon_k^{-\delta}<|1-\theta y_k|_p^{-\delta}. 
\end{equation}

Note that if $|\theta a_k+b_k-2b_{k+1}|_p\not\equiv 0$, then
%$|1-\theta y_k|_p^{1+2\delta}=|1-\theta y_k|_p\cdot |1-\theta y_k|_p^{2\delta}<|1-\theta y_k|_p\epsilon_k^{2\delta}$
\begin{eqnarray*}
%\fl 
%\begin{split}
&|1-\theta y_k|_p^{1+2\delta}=|1-\theta y_k|_p\cdot |1-\theta y_k|_p^{2\delta}<|1-\theta y_k|_p\epsilon_k^{2\delta}
\\
%\fl 
&\leq\frac{|\theta a_k+b_k-2b_{k+1}|_p}{|a_k+\theta b_k|_p}|1-\theta y_k|_p
=|(1+\theta y_{k+2})-B_k|_p
%\\&
\leq\max\left\{|1+\theta y_{k+2}|_p,|B_k|_p\right\}\\
%\fl 
&\leq\max\left\{|1+\theta y_{k+2}|_p,|1-\theta y_k|_p^{3/2-5\delta}\right\}=|1+\theta y_{k+2}|_p.
%\end{split}
\end{eqnarray*}
So $|1+\theta y_{k+2}|_p^{-1}<|1-\theta y_k|_p^{-(1+2\delta)}$.

If $|\theta a_k+b_k-2b_{k+1}|_p\not\equiv 0$ then the second term in (\ref{Ckcase}) satisfies
\begin{equation}
%\fl 
\label{2term}
\left|\frac{B_k(a_k+\theta b_k)}{2\theta (1+\theta y_{k+2})(1-\theta y_k)}\right|_p\leq |1-\theta y_k|_p^{1/2-6\delta}\cdot |1+\theta y_{k+2}|_p^{-1}\leq |1-\theta y_k|_p^{-1/2-8\delta},
\end{equation}
where we have used (\ref{eps2}).  From equation (\ref{Ckcase})
we have
 \begin{eqnarray}
%\begin{split}
%\fl 
&|C_k|_p
\leq \max\left\{\frac{|1/2|_p\cdot |a_{k+2}+\theta b_{k+2}|_p}{|1-\theta y_{k+2}|_p}, \left|\frac{B_k(a_k+\theta b_k)}{2\theta (1+\theta y_{k+2})(1-\theta y_k)}\right|_p, |A_k|_p\right\}
\nn\\
%\fl 
&\leq\max \left\{|1-\theta y_k|_p^{-2\delta}, |1-\theta y_k|_p^{-1/2-8\delta},|1-\theta y_k|_p^{-1/2}\right\}=|1-\theta y_{k}|_p^{-1/2-8\delta}\!\!\!,\ \ \ \ 
\label{upperCk}
%\end{split}
\end{eqnarray}
where we have used (\ref{1-yk+2}), (\ref{2term}) and the first part of the theorem.  From (\ref{k+2relk}) we have
$|1+\theta y_{k+2}|_p\leq |1-\theta y_k|_p^{1-2\delta}$. 
%Taking the reciprocal of the above inequality and raising both sides to the power $\frac{(1/2+8\delta)}{1-2\delta}$, we get
So for sufficiently small $\delta$
$$
|C_k|_p\le|1-\theta y_k|_p^{-(1/2+8\delta)}%\leq |1+\theta y_{k+2}|_p^{\frac{-(1/2+8\delta)}{(1-2\delta)}}
\leq |1+\theta y_{k+2}|_p^{-1/2-10\delta}\leq|1+\theta y_{k+2}|_p^{-2/3-2\delta}.
$$

Now if $|\theta a_k+b_k-2b_{k+1}|_p\equiv 0$, then the upper bound on the second term in (\ref{Ckcase}) is
$$
\left|\frac{B_k(a_k+\theta b_k)}{2\theta (1+\theta y_{k+2})(1-\theta y_k)}\right|_p=\frac{|1+\theta y_{k+2}|_p\cdot|a_k+\theta b_k|_p}{|2|_p \cdot|1+\theta y_{k+2}|_p\cdot|1-\theta y_k|_p}%\nonumber\\
\leq|1-\theta y_k|_p^{-(1+\delta)}. 
$$
Consequently, 
\begin{eqnarray*}
%\begin{split}
|C_k|_p&\leq& \max\left\{\frac{|1/2|_p\cdot |a_{k+2}+\theta b_{k+2}|_p}{|1-\theta y_{k+2}|_p}, \left|\frac{B_k(a_k+\theta b_k)}{2\theta (1+\theta y_{k+2})(1-\theta y_k)}\right|_p, |A_k|_p\right\}\\
&\leq&\max\{|1-\theta y_k|_p^{-2\delta}, |1-\theta y_k|_p^{-(1+\delta)}, |1-\theta y_k|_p^{-1/2}\}=|1-\theta y_k|_p^{-(1+\delta)}.
%\end{split}
\end{eqnarray*}
Since $|1+\theta y_{k+2}|_p=|B_k|_p\leq |1-\theta y_k|_p^{3/2-5\delta}$, it yields that $|1-\theta y_k|_p^{-(1+\delta)}\leq |1+\theta y_{k+2}|_p^{\frac{-(1+\delta)}{3/2-5\delta}}\leq |1+\theta y_{k+2}|_p^{-2/3-2\delta}$. Hence, $|C_k|_p\leq  |1+\theta y_{k+2}|_p^{-2/3-2\delta}$ which proves the last part of the theorem for the non-Archimedean absolute value.
% ($p<\infty$).\\

%%%%%%%%%%%%%%%%%%% Archimedean absolute value%%%%%%%%%%%%%%

%\pagebreak

%%%%%

Estimates for the Archimedean case ($p=\infty$) are similar to the above.  Here we will derive the estimate for $|A_k|_p$ only. %As in the non-Archimedean absolute value case, we could rewrite $A_k$ as $A_k=\frac{\textstyle a_k-\theta b_k}{\textstyle 2(1+\theta y_k)}-y_{k-1}$. Applying the Archimedean absolute value for both sides of the equation and using the triangle inequality, we get
 Since
$2=|2|_\infty\leq |1-\theta y_k|_\infty+|1+\theta y_k|_\infty
<\epsilon_k+|1+\theta y_k|_\infty
<1+|1+\theta y_k|_\infty$, we have
$|1+\theta y_k|_\infty^{-1}<1$.
%\begin{equation}
%\label{arch1+yk}
%\frac{1}{|1+\theta y_k|_p}< 1.
%\end{equation}
So equation (\ref{Akforms}) gives
\begin{eqnarray*}
|A_k|_\infty&\leq& \frac{|1/2|_\infty\cdot |a_k-\theta b_k|_\infty}{|1+\theta y_k|_\infty}+|y_{k-1}|_\infty
\leq\frac{1}{10}\epsilon_k^{-\delta}\cdot 1+|1-\theta y_k|_\infty^{-1/2}\\
&\leq&\frac{1}{10}|1-\theta y_k|_\infty^{-\delta}+|1-\theta y_k|_\infty^{-1/2}
\leq\frac{11}{10}|1-\theta y_k|_\infty^{-1/2},
\end{eqnarray*}
for sufficiently small $\delta$, which proves the first part of the theorem for $p=\infty$.
\hfill $\Box$
%This proves the theorem for Archimedean absolute value and the proof is completed for Theorem \ref{SC}.    $\hspace{13.5cm}\square$ 
 %\end{description}

%\bibliographystyle{plain}
%\bibliography{references}

\vskip 5mm

\noindent{\large\bf Acknowledgements}\vskip 2mm

\noindent
The first author thanks Sultan Qaboos University for supporting her through grant IG/SCI/DOMAS/12/01.  The second author gratefully acknowledges support from EPSRC grants EP/C54319X/2 and EP/I013334/1.

%\section*{References}

\end{document}